\documentclass[iop,twocolumn,twocolappendix,appendixfloats]{emulateapj}
\usepackage[english]{babel}
\usepackage[utf8x]{inputenc}
\usepackage{amsmath}
\usepackage{graphicx}
\usepackage{natbib}
\pdfoutput=1
\usepackage{color}
\usepackage[usenames,dvipsnames]{xcolor}
\usepackage{mathtools}
\usepackage{comment}
\usepackage{threeparttable}
\usepackage{multirow}

\newcommand\appsection{
   \setcounter{section}{0}
   \renewcommand*{\thesection}{\Alph{section}}
}

\newcommand{\hii}{H\,{\sc ii}}

\newcommand{\cii}{C\,{\sc ii}}
\newcommand{\oi}{O\,{\sc i}}
\newcommand{\nii}{N\,{\sc ii}}

\newcommand{\mum}{\ensuremath{\,\mu\mbox{m}}}

\newcommand{\degree}{$^{\circ}$}

\bibpunct[, ]{(}{)}{;}{a}{,}{,}

\begin{document}

\title{The survey of lines in M\,31 (SLIM): The Drivers of the [\cii]/TIR Variation}

\author{Maria J. Kapala\altaffilmark{1,2}, Brent Groves\altaffilmark{3}, Karin Sandstrom\altaffilmark{4}, Thomas Jarrett\altaffilmark{1}, Elisabete da Cunha\altaffilmark{3}, Kevin Croxall\altaffilmark{5,6}, Julianne Dalcanton\altaffilmark{7}, Bruce Draine\altaffilmark{8}, Simon Glover\altaffilmark{9}, Eva Schinnerer\altaffilmark{2}}
\altaffiltext{1}{University of Cape Town, Cape Town, Republic of South Africa}
\altaffiltext{2}{Max Planck Institute for Astronomy, Heidelberg, Germany}
\altaffiltext{3}{Research School of Astronomy \& Astrophysics, Australian National University, Canberra, Australia}
\altaffiltext{4}{Center for Astrophysics and Space Sciences, University of California, San Diego, USA}
\altaffiltext{5}{Department of Astronomy, Ohio State University, 140 West 18th Avenue, Columbus, USA}
\altaffiltext{6}{Illumination Works LLC, 5650 Blazer Parkway, Dublin, USA}
\altaffiltext{7}{University of Washington, Seattle, USA}
\altaffiltext{8}{108 Peyton Hall, Princeton University, Princeton, USA}
\altaffiltext{9}{Zentrum f\"ur Astronomie der Universit\"at Heidelberg, Institut f\"ur Theoretische Astrophysik, Albert-Ueberle-Str. 2, 69120 Heidelberg, Germany}

%===================================================================

\begin{abstract}
The ratio of the [\cii] 158$\,\mu$m emission line over the total infrared emission (TIR) is often used as a proxy for the photoelectric (PE) heating efficiency ($\epsilon_{\rm PE}$) of the far-ultraviolet (FUV) photons absorbed by dust in the interstellar medium. 
In the nearby galaxy M\,31, we measure a strong radial variation of [\cii]/TIR  that we rule out as being due to an intrinsic variation in $\epsilon_{\rm PE}$.
[\cii]/TIR fails as a proxy for $\epsilon_{\rm PE}$, because the TIR measures all dust heating, not just the contribution from FUV photons capable of ejecting electrons from dust grains. 
Using extensive multiwavelength coverage from the FUV to far-infrared (FIR), we infer the attenuated FUV emission ($\rm UV_{att}$), and the total attenuated flux ($\rm TOT_{att}$).
We find [\cii]/TIR to be strongly correlated with $\rm UV_{att}$/$\rm TOT_{att}$,  indicating that, in M\,31 at least, one of the dominant drivers for [\cii]/TIR variation is the relative hardness of the absorbed stellar radiation field.
We define $\rm{ \epsilon_{PE}^{UV}}$,  [\cii]/$\rm{ UV_{att}}$ which should be more closely related to the actual PE efficiency, which we find to be essentially constant ($1.85 \pm 0.8 \%$) in all explored fields in M\,31. 
This suggests that part of the observed variation of [\cii]/TIR  in other galaxies is likely due to a change in the relative hardness of the absorbed stellar radiation field, caused by a combination of variations in the stellar population, dust opacity and galaxy metallicity, although PE efficiency may also vary across a wider range of environments.
\end{abstract}

%===================================================================

\section{Introduction}
\label{sec:bulge_intro}

The balance between radiative heating and cooling dictates the multiphase structure of the ISM, which in turn controls the process of star formation (SF). Understanding the energy balance of the interstellar medium (ISM) is therefore an essential step towards understanding the processes which shape the evolution of galaxies.
The photoelectric (PE) effect on dust grains is the dominant heating mechanism in the neutral ISM, in particular the diffuse gas and dense photodissociation regions (PDRs) surrounding \hii\ regions \citep{Weingartner2001, Wolfire2003}. 
While interstellar dust absorbs both ultraviolet (UV) and visible photons, electrons are ejected from dust grains
only by photons with energy exceeding the energy threshold, ${\rm E_{thr}}$, the sum of the work function $W$ and the Coulomb potential $\phi$ \citep[i.e. ${\rm E_{thr}} \gtrsim 6$\,eV, equivalent to $\lambda = 2067$\,\AA;][]{Draine1978,Tielens1985}. Any excess energy is carried by the kinetic energy of the photoejected electron, heating the surrounding ISM.
There has been much analytic work done on the topic of the PE effect \citep{Watson1972, Draine1978, Tielens1985, Bakes1994, Weingartner2001} since \cite{Spitzer1948} first considered it as an important heating mechanism.

The probability that absorption of a photon will result in a photoelectron is a function of the energy of the incident photon $h\nu$, the composition and size of the grain, and the charge of the grain before the event. 
Following \citet{Tielens2008}, we define the theoretical photoelectric heating efficiency as:
\begin{equation}
{\rm \epsilon_{PE}^{th} = \frac{\Gamma_{heat}}{\Gamma_{abs}},}
\label{eq:theo}
\end{equation}
where $\Gamma_{heat}$ and $\Gamma_{abs}$ indicate the photoelectric heating rate and the grain FUV absorption rate, respectively. 
Estimates of the photoelectric heating efficiency typically lie in the range ${\rm \epsilon_{PE}^{th}} \sim 0.1 - 5\%$ \citep{Bakes1994,Weingartner2001}. Its value within this range depends on the relative effectiveness of grain photoionization and recombination, and hence on the parameter ${\rm \gamma=G_0 T^{1/2}/n_e}$ (where G$_0$ is the interstellar radiation field strength integrated between 6 and 13.6\,eV in units of the \citet{Habing} field, $T$ is the gas temperature and n$_{e}$ is the electron density).

%--------------------------------------------

The [\cii] 158\,\mum\ emission line is the dominant coolant in the bulk of the neutral interstellar medium  \citep{Hollenbach1991}, with the exception of the high density and high temperature regime where [\oi] 63\,\mum\ emission can become a dominant coolant \citep[e.g.][]{ Kaufman1999}. [\cii] is one of the brightest emission lines in most galaxies. It is a forbidden fine-structure line arising from singly ionized carbon C$^+$ (${\rm ^2P_{3/2} \rightarrow}$ ${\rm ^2P_{1/2}}$).
Due to its wavelength and brightness, it is  detectable out to high redshifts \citep{Walter2009, Maiolino2015, Capak2015, Willott2015} with millimeter telescopes.

One of the driving reasons to examine the [\cii] line is its potential to trace star formation rates (SFRs) out to high redshift \citep{Carilli2013}. 
Empirically, the correlation between [\cii] and SFR
is close to one-to-one on $\sim$\,kpc scales in nearby galaxies 
\citep{deLooze2011, deLooze2014, Herrera2015}.
Theoretically, the basis for [\cii] as a SFR indicator relies upon it being the dominant coolant of a neutral ISM heated by massive stars, associated with
recent star formation ($\lesssim$ few $10^7$\,yr), through the PE effect.

Unfortunately, there are potential issues with [\cii] as a SFR tracer. First, C$^+$ has a lower ionization potential than hydrogen at 11.2\,eV, and thus
[\cii] emission can arise from different interstellar medium phases: the cold and warm neutral medium (CNM and WNM, respectively), and the warm ionized medium (WIM). Therefore, there might be a different proportionality between [\cii] and SFR in different phases if other coolants (e.g. the [\oi] 63\,\mum\ emission line) or heat sources (e.g., photons from older stars, cosmic-rays, or X-rays from an AGN) are in play. 
Second, if the PE heating efficiency is variable or evolves with redshift, the use of [\cii] as a SFR tracer may be complicated.
Understanding the efficiency of PE heating is therefore crucial for establishing [\cii] as a robust star formation indicator.

Unfortunately,  ${\rm \epsilon_{PE}^{th}}$ cannot be directly observationally measured.
Instead, the most commonly used  approximation of the photoelectric heating efficiency is based on observable properties:
\begin{equation}
{\rm  \epsilon_{\rm PE}^{\rm dust} = \frac{\eta_{[CII]} \Gamma_{gas}(6-13.6\,eV)}{\Gamma_{dust}(0-13.6\,eV)} \simeq \frac{[CII]}{TIR}}
\label{eq:PEeff_dust}
\end{equation}
where the $\Gamma$'s are the heating rates of the gas and the dust, and ${\rm \eta_{[CII]}}$ is the fraction of the total gas cooling that is due to [\cii] \citep{Mochizuki2004}. This approximation works under the assumption that the ISM is in steady state (heating equals cooling),
and that the [\cii] emission dominates neutral gas cooling, so that $\rm \eta_{[CII]}\simeq 1.$ 
The latter assumption is reasonable in low density gas \citep[e.g.][]{Wolfire2003}, but in the high density and temperature regime [\oi] 63\,\mum\ emission can become a dominant coolant, and so in some studies, [\cii]+[\oi] is used in place of [\cii] as the numerator in Equation~\ref{eq:PEeff_dust} \citep[see e.g.][]{Croxall2012, Cormier2015}.
The dust is assumed to be in thermal equilibrium, so that its heating rate can be derived by integrating its continuum emission. Since this emission is almost exclusively located in the infrared, it is generally sufficient to compute the total infrared emission of the dust, TIR, as an approximation of $\rm \Gamma_{dust}$.

As mentioned above, neutral gas heating arises predominantly due to collisions with photoelectrons ejected by FUV photons (6--13.6\,eV), while dust heating arises due to absorption of photons of all energies (mostly 0.01--13.6\,eV). 
Therefore, we emphasize here that the [\cii]/TIR approximation diverges from the theoretical definition (Eq.~\ref{eq:theo}), since soft optical photons which have energies below the work threshold heat the dust but do not contribute to the PE effect.
Thus, $\epsilon_{\rm PE}^{\rm dust}$ is sensitive to the ratio of the soft photons
(transferring energy only to dust) to FUV photons (transferring energy to both dust and gas). A similar issue has been raised by \citet{Okada2013}, for deriving $G_0$ from the TIR , where they show that only $\lesssim 25\%$ of the total stellar radiation emitted by a B9 star is in the 6--13.6\,eV energy range, based on the \citet{Castelli2004} stellar atmosphere model.

Nevertheless,  [\cii]/TIR has been extensively used as a proxy for PE heating efficiency in the last 25 years.
Studying this ratio led to the discovery of the so called ``[\cii] deficit'', defined as [\cii]/TIR declining with warmer dust color \citep[a proxy for dust temperature;][] {Malhotra2001}, or as simply  [\cii]/TIR values below $10^{-3}$ \citep{Helou2001}. 
The observed ``[\cii] deficit'', typically treated as a reduction of the PE efficiency, has been explained to arise by several possible mechanisms, 
including an increase in the mean grain charge (and Coulomb potential $\phi$), and therefore a higher E$_{\rm thr}$ for photoelectron ejection 
\citep{Malhotra2001, Croxall2012}, the depletion of 
polycyclic aromatic hydrocarbons (PAHs; very efficient contributors of photoelectrons) relative to larger dust grains \citep{Helou2001}, 
or a significant fraction of the TIR arising from dusty \hii\ regions, where [\cii] is not the dominant coolant \citep{Luhman2003, GraciaCarpio2011}.

Many recent studies have used well-resolved galaxies in the local universe to study the environmental dependance of [\cii]/TIR, a PE heating efficiency proxy, including the impact of metallicity, dust grain size distribution, radiation fields etc. Most of the studies found large [\cii]/TIR variations with these parameters. 
For example, \cite{Cormier2012,Cormier2015} found higher [\cii]/TIR in lower-metallicity galaxies.
\citet{Rubin2009}  studied systematic variations in [\cii]/TIR in the LMC, and found the highest [\cii]/TIR in the diffuse medium, while the lowest ratios coincide with bright star-forming regions ($\sim$\,1.4 times lower).  However, the bulk of the [\cii]/TIR values are scattered around 0.005. 
\citet{Smith2016} showed that in a diverse sample of nearby galaxies, there are trends of decreasing [\cii]/TIR with increasing metallicity and with increasing star formation rate surface density.
Other studies have shown that PAHs are the most efficient dust constituents in the PE heating mechanism  \citep{Rubin2009}, and [\cii]/TIR decreases with increasing ionized fraction of PAHs \citep{Okada2013}. As such, \citet{Helou2001} and \citet{Croxall2012} suggested that [\cii]/PAH is a more direct measure of the heating efficiency in the ISM than [\cii]/TIR.
Finally, \citet{Mizutani2004} found decreasing [\cii]/TIR with increasing UV radiation field.
The point is that [\cii]/TIR depends strongly on the local galactic environment. 
Many of the above dependencies are entangled, making any conclusive statements difficult.

Using the Survey of Lines In M\,31 (SLIM), \citet{Kapala2015} showed that the [\cii]/TIR ratio  varies dramatically with radius in M\,31's disk, spanning a factor of 3 between 6.9\,kpc and 16.0\,kpc; see top panel Figure~\ref{fig:UVatt_TOTatt_radius}. 
This revelation is surprising, as \citet{Kapala2015} also showed that [\cii] correlates well with SFR even on small scales ($\sim$\,50\,pc). 
How can the [\cii]-SFR relation hold even when the [\cii]/TIR proxy of PE efficiency appears to change?

\citet{Kapala2015} presented two possible scenarios that could cause the variation in [\cii]/TIR without directly changing the PE efficiency (therefore allowing the [\cii]-SFR relationship to hold): (1) a change in the hardness of the integrated stellar spectrum, changing the relative abundance of photoelectric photons to dust heating photons
and/or (2) a change in the opacity due to varying gas surface density or dust-to-gas ratio \citep[e.g. allowing photons at longer wavelengths to escape preferentially for lower dust-to-gas ratio or low overall column density of dust;][]{Israel1996}.

The distinction between ${\rm \epsilon_{PE}^{th}}$ and [\cii]/TIR raises an issue.
A significant fraction of stellar radiation is absorbed by dust in the wavelength range softer than UV. \citet{Viaene2016} find that on average $\sim 44\%$ of the dust luminosity comes from absorption of non-UV photons, based on a sample of 239 galaxies. Their conservative UV definition spans to a longer wavelength ($\lambda = 3650$\,\AA), therefore integrating  to a shorter  wavelength upper limit of  $\lambda = 2067$\,\AA (as we use here), would lead to an even larger percentage.
Therefore, a large fraction of absorbed stellar energy contributes only to dust heating, but is not energetic enough to heat the gas. 
 This leads to a degeneracy: different relative hard (UV) to soft (optical) photon dust heating contributions can result in the same TIR value.

With this paper we investigate the behaviour of the photoelectric efficiency in M\,31 using a more carefully motivated tracer for $\epsilon_{\rm PE}$ than [\cii]/TIR. We  constrain how much of the observed [\cii]/TIR variation can be tied to PE efficiency, and  test other explanations for why [\cii]/TIR might vary.

This paper is organized as follows: in Section~\ref{sec:data}, we describe the data and processing. 
In Section~\ref{sec:int_stel_em}, we apply {\tt MAGPHYS}, an SED modelling tool to study the energy balance in the ISM.
In  Section~\ref{sec:results}, we determine the PE efficiency in the SLIM fields using different approximations, and assess  how the underlying assumptions affect measurements.
In Section~\ref{sec:disc}, we discuss our results and other possible factors potentially contributing to the variation of the [\cii]/TIR ratio, and the global impact of our results on observational estimates of PE heating efficiency. We summarize our findings in Section~\ref{sec:concl}.

\begin{figure*}[!ht]
\begin{center}
\includegraphics[width=1.\textwidth]{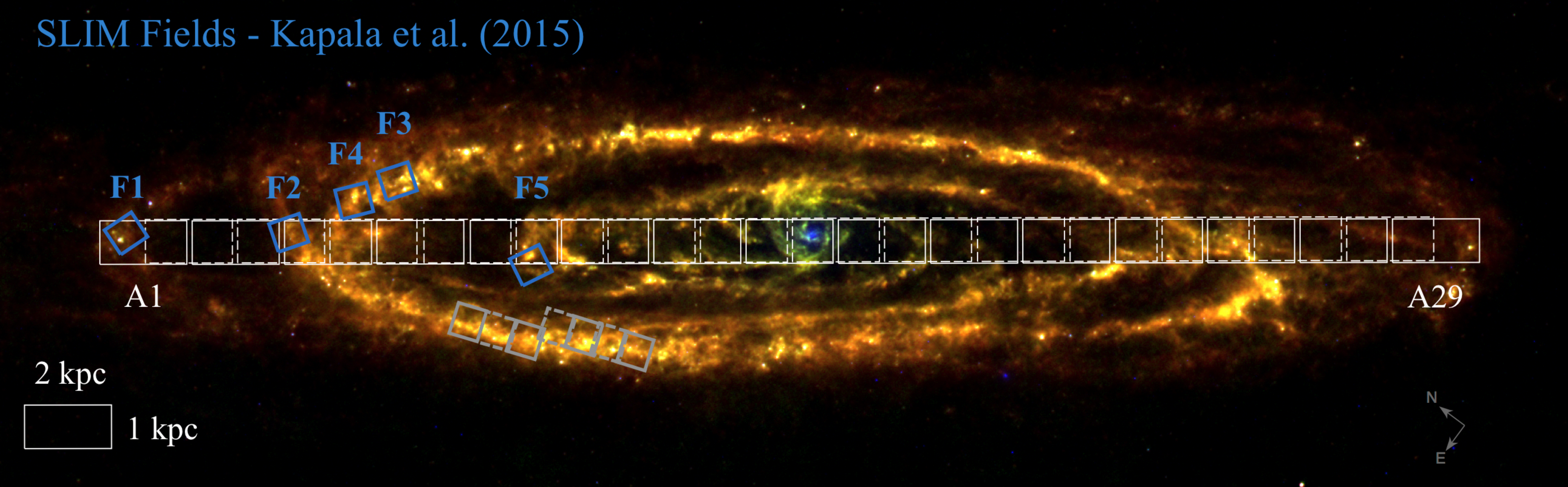}
\caption{Three color dust emission image (red - SPIRE 250\,\mum, green - PACS 160\,\mum\ and blue - MIPS 24\,\mum). The white 1$\times$2\,kpc apertures (solid and dashed squares) indicate where the integrated SED modeling with {\tt MAGPHYS} along the major axis of M\,31 has been performed. The SLIM survey fields are overlaid in blue. 
The coverage of archival {\em ISO} observations from \citet{Rodriguez2006} are shown as gray boxes.  }
\label{fig:m31_1kpc_ap}
\end{center}
\end{figure*}

%===================================================================

\section{Data}
\label{sec:data}

The key data in this paper are maps of [\cii] emission combined with multi-wavelength imaging of M\,31 spanning from far-UV to far-IR wavelengths. 
Henceforth, we assume the following parameters for M\,31; nucleus position $RA=00^h42^m44.^s35$, $Dec=+ 41^{\circ}16\arcmin08\farcs60$ (J2000)\footnote{Based on NED data and references therein}; inclination 70$^{\circ}$ and position angle 43.2$^{\circ}$ \citep{Dalcanton2012}; Distance $780 \pm 40$\,kpc \citep{Stanek1998}.

We mapped the [\cii] 158\,\mum\ emission line in five $\sim$\,700\,$\times$\,700\,pc fields \citep[henceforth SLIM fields;][]{Kapala2015} with \emph{Herschel} PACS at the resolution $\sim$\,50\,pc, FWHM $\sim$ 11\farcs4. The coordinates of the SLIM fields are given in Table~\ref{tab:obs}. 
For details on the integral field spectroscopic data reduction see \citet{Kapala2015}.

We also use \emph{Herschel} and \emph{Spitzer} infrared (IR) photometry. Observations of M\,31 were done with the Multiband Imaging Photometer \citep[MIPS;][]{Rieke2004} on board of the {\em Spitzer} Space Telescope \citep{Werner2004} by \citet{Gordon2006} and the InfraRed Array Camera \citep[IRAC;][]{Fazio2004} by \citet{Barmby}.  
At the longer wavelengths, M\,31 was observed with the {\em Herschel} Space Observatory (PI O. Krause) using the PACS and SPIRE instruments \citep[Spectral and Photometric Imaging Receiver;][]{Griffin2010}.  \emph{Herschel} observations and processing details can be found in \citet{Groves2012} and \citet{Draine2014}. 

Details about all photometric bands used in this paper, including effective wavelength, FWHM and 1-$\sigma$ sensitivity are presented in Table~\ref{tab:data_sum}.

We use far-UV data from the \emph{Galaxy Evolution Explorer} (\emph{GALEX}). 
FUV ($1350-1750$\,\AA) and NUV ($1750-2800$\,\AA) observations of M\,31 \citep{Thilker2005} were obtained as part of the \emph{GALEX} Nearby Galaxy Survey \citep[NGS;][]{Bianchi2003,Bianchi2005}.

For the optical photometry we used the Sloan Digital Sky Survey \citep[SDSS;][]{York2000} data, which covered the entirety of the disk of M\,31 in a contiguous strip using the $ugriz$ filters. 
Here we use the $ugriz$ mosaics of \citet{Tempel2011}. Details of the reduction of the \emph{SDSS} images and the important sky removal are described in \citet{Tempel2011} and Tempel et al. (in prep.). The units are nanomaggies per pixel, where nanomaggie\footnote{$www.sdss3.org/dr8/algorithms/magnitudes.php\#nmgy$} is a linear unit of flux density equal to $10^{-9}$\,maggies. A star of brightness 1\,nanomaggie has an AB magnitude of 22.5 in any band, and a flux density equal to $3.631\times10^{-6}$\,Jansky. 

For the comparison with other studies, we use H$\alpha$ maps from the optical IFS from Calar Alto 3.5\,m telescope with the PMAS instrument in PPAK mode with the V300 grating \citep{Roth2005,Kelz2006}. For all of the observing and reduction details see \citet{Kapala2015}. 

\begin{table}[h!]\centering
\caption{Photometric data properties}
\begin{threeparttable}
\begin{tabular}{|c|c|c|c|c|}
\hline
Band & $\lambda_{eff}$ & FWHM    & Pix size& 1-$\sigma$   \\
        &            &           &         & noise  \\
\hline
FUV     & 1516\,\AA  & 4\farcs0  & 1\arcsec   &   6.6$\times 10^{19}$\,erg\,s$^{-1}$\,cm$^{-2}$ \\
NUV     & 2267\,\AA  & 5\farcs6  & 1\arcsec   &   2.8$\times 10^{19}$\,erg\,s$^{-1}$\,cm$^{-2}$  \\
SDSS u  & 3543\,\AA  & 1\arcsec  & 0\farcs396 &  0.21 MJy\,sr$^{-1}$  \\
SDSS g  & 4770\,\AA  & 1\arcsec  & 0\farcs396 &  0.34 MJy\,sr$^{-1}$  \\
SDSS r  & 6231\,\AA  & 1\arcsec  & 0\farcs396 &  0.43 MJy\,sr$^{-1}$  \\
SDSS i  & 7625\,\AA  & 1\arcsec  & 0\farcs396 &  0.48 MJy\,sr$^{-1}$  \\
SDSS z  & 9134\,\AA  & 1\arcsec  & 0\farcs396 &  1.46 MJy\,sr$^{-1}$  \\
IRAC 1  & 3.6\,\mum\ & 1\farcs66 & 1\farcs2   &   21.2 mag arcsec$^{-2}$ \\
IRAC 2  & 4.5\,\mum\ & 1\farcs72 & 1\farcs2   &   20.9 mag arcsec$^{-2}$ \\
IRAC 3  & 5.8\,\mum\ & 1\farcs88 & 1\farcs2   &   19.0 mag arcsec$^{-2}$ \\
IRAC 4  & 8.0\,\mum\ & 1\farcs98 & 1\farcs2   &   19.6 mag arcsec$^{-2}$\\
MIPS    & 24\,\mum\  & 6\farcs5  & 2\farcs55  &  0.12 MJy\,sr$^{-1}$ \\
        & 70\,\mum\  & 18\arcsec & 9\farcs98  &  0.21 MJy\,sr$^{-1}$ \\
PACS\tnote{a}  & 70\,\mum\  & 5\farcs6  & 1\arcsec   & 3.17 MJy\,sr$^{-1}$ \\ 
        & 100\,\mum\ & 6\farcs8  & 1\arcsec   & 3.23 MJy\,sr$^{-1}$  \\
        & 160\,\mum\ & 11\farcs4 & 1\arcsec   & 2.29 MJy\,sr$^{-1}$  \\
SPIRE\tnote{a} & 250\,\mum\ & 17\farcs6 & 6\arcsec   & 0.68 MJy\,sr$^{-1}$  \\
        & 350\,\mum\ & 24\farcs9 & 10\arcsec  & 0.46 MJy\,sr$^{-1}$  \\
        & 500\,\mum\ & 36\farcs3 & 14\arcsec  & 0.20 MJy\,sr$^{-1}$  \\
\hline
\end{tabular}

\begin{tablenotes}
{\footnotesize
\item[a] \emph{Herschel} Observer's Manuals \\ 
${\rm herschel.esac.esa.int/Docs/PACS/html/pacs\_om.html}$, \\ ${\rm 
herschel.esac.esa.int/Docs/SPIRE/html/spire\_om.html}$  \\ for 20\arcsec s$^{−1}$ scans. }
\end{tablenotes}

\end{threeparttable}
\label{tab:data_sum}
\end{table}

%-------------------------------------------

\subsection{Processing of the data}
\label{sec:proc_data}

We use {\it Herschel} and {\it Spitzer} images that have been processed following a general method described in \citet{Aniano2012}, and details specific to M\,31 data are reported in \citet{Draine2014}. 

We summarize the most important processing steps. A ``tilted plane'' best-model background has been fitted to each IR image (except IRAC 5.8 and 8\,\mum, where more complex models were necessary) in regions outside the galaxy. 
A simple constant value background model has been used in case of {\it GALEX} \citep[0.000929\,MJy\,sr$^{-1}$ for FUV, and 0.004712\,MJy\,sr$^{-1}$ for NUV][]{Thilker2005} and \emph{SDSS} data. 
Corresponding background models have been subtracted from the images. Then, we convolve all data with the {\tt convolve\_fft} package \citep{Astropy2013} to match the MIPS 160\,\mum\ resolution using dedicated kernels from \citet{Aniano2011}. 
The MIPS 160\,\mum\ resolution allows us to use all photometric bands in our analysis, including the longest wavelength band, i.e. SPIRE 500\,\mum. 
Finally, we resample the convolved images to the same pixel grid of 18\arcsec by 18\arcsec pixel size, which is the lower limit of the Nyquist sampling of the largest beam, i.e. $0.5 \times $\,FWHM${\rm _{M160\mu m}}$. We report 1-$\sigma$ noise and other filter properties of interest in  Table~\ref{tab:data_sum}.

%-------------------------------------------

\begin{table}[h!]\centering
\caption{Coordinates of the SLIM Field centers. }
\begin{threeparttable}
\begin{tabular}{|c|c|c|c|c|}
\hline
F & R.A.     & Dec.     & P.A.        & Observation\tnote{a}  \\
  & (J2000)  & (J2000)  &[$^{\circ}$] & ID \\
\hline
1     & $00^h46^m29.^s17$ & $+42^{\circ}11\arcmin30\farcs89$ & 70.7  & 1342236285 \\
2     & $00^h45^m34.^s79$ & $+41^{\circ}58\arcmin28\farcs54$ & 55.7  & 1342238390  \\
3     & $00^h44^m36.^s49$ & $+41^{\circ}52\arcmin54\farcs21$ & 55.0  & 1342238391  \\
4     & $00^h44^m59.^s26$ & $+41^{\circ}55\arcmin10\farcs47$ & 51.0  & 1342238726 \\
5     & $00^h44^m28.^s76$ & $+41^{\circ}36\arcmin58\farcs91$ & 63.0  & 1342237597 \\
\hline
\end{tabular}

\begin{tablenotes}
{\footnotesize
\item[a] [\cii] \& [\oi] observations acquired between 28$^{\rm th}$ Feb and 1$^{\rm st}$ Mar, 2012}
\end{tablenotes}

\end{threeparttable}
\label{tab:obs}
\end{table}

%===================================================================

\section{The intrinsic stellar emission of M\,31}
\label{sec:int_stel_em}

To infer the intrinsic UV--optical spectra of the regions we assume that the energy is conserved between the UV--optical absorption and IR emission and use spectral energy distribution (SED) modelling of the integrated photometry.\footnote{We note that on small scales where the influence of individual star forming regions can be resolved, energy may not be conserved within a pixel. However, since we work at $\sim$\,kpc scales, we expect this to be a good approximation. \citet{Parravano}  showed that it is stars within $\sim$\,500\,pc that create the radiation field near the Sun.}

We infer the intrinsic stellar UV-optical continuum to determine the spectrum absorbed by dust and the fraction going into PE heating.

%-----------------------------------------------------------------

\subsection{SED fitting}
\label{sec:magphys}

The SED fitting code {\tt MAGPHYS} \citep{daCunha2008} compares the photometric observations of galaxies with computed photometry from libraries of stellar population synthesis model spectra and model IR SEDs.
The template stellar population spectra  are derived using the \citet{Bruzual2003} stellar synthesis code.
The stellar libraries provide high resolution spectra from the rest frame UV to IR and include a range of star formation histories, including bursts,
assuming a \citet{Chabrier2003} initial mass function (IMF). The final stellar library spectra are attenuated by dust over a range of $A_{V}$'s assuming
a two-component model \citep{Charlot2000}, that takes into account the effect of the higher attenuation of young ($\lesssim10^7$\,yr) stars in 
the natal clouds ($\tau_{\lambda}^{BC}$, optical depth in birth clouds), in addition to the attenuation of all stars by dust in the ambient ISM ($\tau_{\lambda}^{ISM}$).

The library of dust emission spectra is calculated from a four-component model created specifically for {\tt MAGPHYS} by \citet{daCunha2008}. 
The main contributors to this model are PAHs (emitting spectral features between 3 and 20\,\mum),
stochastically heated very small grains ($a<0.005$\,\mum, emitting mainly at mid-IR wavelengths $\lesssim 40$\,\mum), and two emitting components of cold and warm large grains in thermal equilibrium (emitting mainly at far-IR wavelengths). 

\begin{figure}[!ht]
\begin{center}
\includegraphics[width=.5\textwidth]{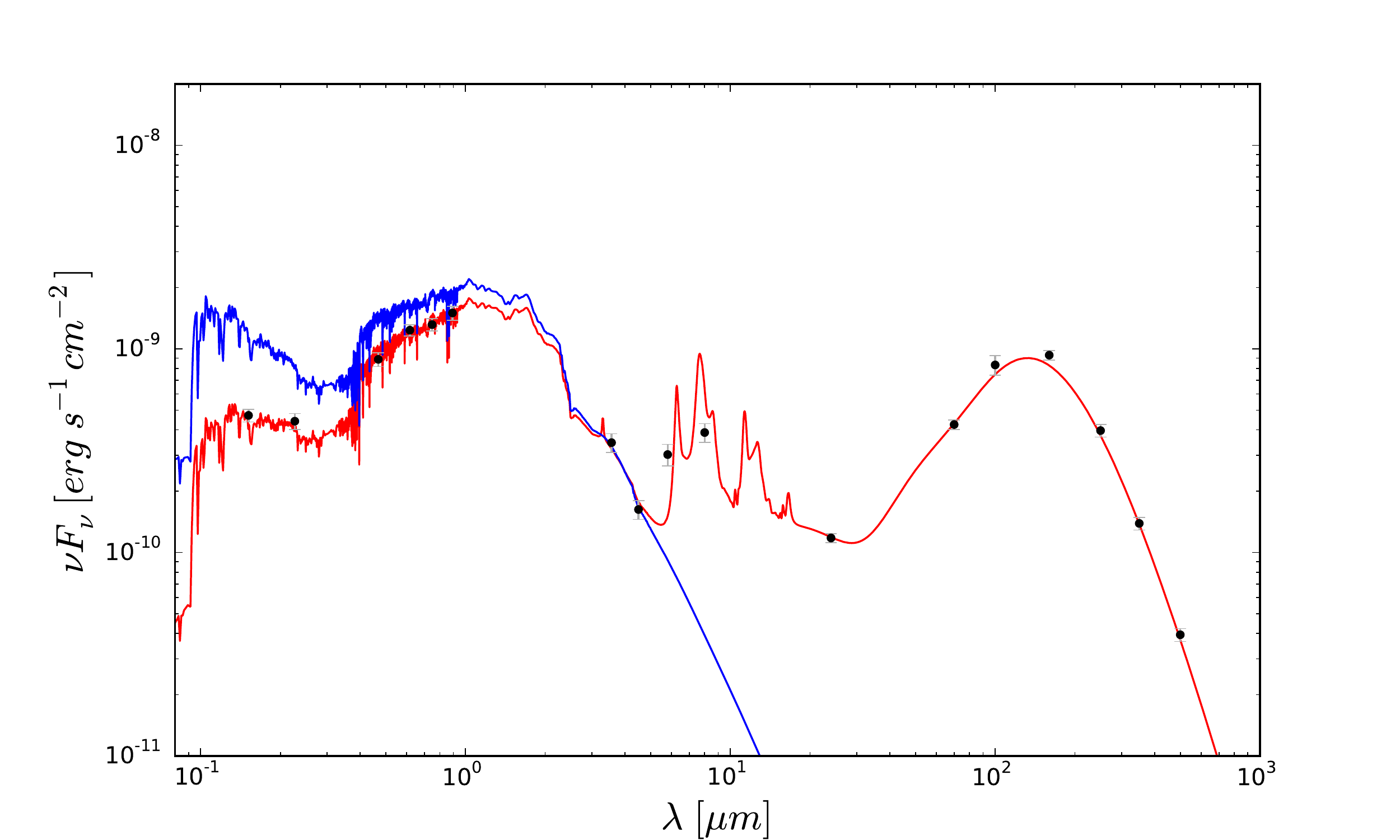}
\caption{The SED of SLIM field 3 in M\,31, measured at a distance of 11.3\,kpc from the center. The photometric measurements used for the fit: GALEX FUV and NUV, SDSS g, r, i and z, {\em Spitzer} IRAC and MIPS\,24\,\mum, and all {\em Herschel} bands except SPIRE\,500\,\mum.
The red curve is the best fit {\tt MAGPHYS} spectrum to the data (stellar and dust emission), while the blue curve is the inferred unreddened stellar spectrum.}
\label{fig:sed_f3}
\end{center}
\end{figure}

The strength of {\tt MAGPHYS} lies in the consistent modelling of the stellar and dust emission from the FUV to the far-IR via an energy balance technique: the angle-averaged amount of starlight absorbed by dust in the FUV to near-IR is computed using realistic star formation histories and dust attenuation prescription, and this energy is then re-radiated in the mid- to far-IR range using empirically-calibrated dust emission components. MAGPHYS uses a two-component ISM geometry to account for the attenuation of starlight from stellar birth clouds and from the diffuse ISM, following \citet{Charlot2000}. The resulting attenuation curve is flatter than a typical extinction law because it accounts for a geometry where the stars and dust clouds are mixed, and includes the scattering of short wavelength stellar emission back into the line of sight. Moreover, {\tt MAGPHYS} allows for a 15\% variation between the total dust emission and the attenuated stellar emission to account for any geometric discrepancies not accounted for in a simple isotropic model.
Finally {\tt MAGPHYS} uses a Bayesian approach, and calculates the likelihood distribution of each physical parameter 
by evaluating how well each SED in the model grid accounts for the observed properties of the galaxy \citep[for details on the assumed priors see][]{daCunha2008}.

%-----------------------------------------------------------------

\subsection{Quantifying attenuated stellar emission}
\label{sec:attenuation}

We use the intrinsic and reddened stellar spectra fitted with {\tt MAGPHYS} to determine the stellar energy contribution to the ISM heating. We present an example SED fit to SLIM field 3 in Figure~\ref{fig:sed_f3}.
The attenuated stellar emission  is the difference between the intrinsic (unreddened; blue curve in Figure~\ref{fig:sed_f3}) and reddened model
spectrum (red curve in Figure~\ref{fig:sed_f3}) that best fit the observed data.

\begin{figure}[!ht]
\begin{center}
\includegraphics[width=.5\textwidth]{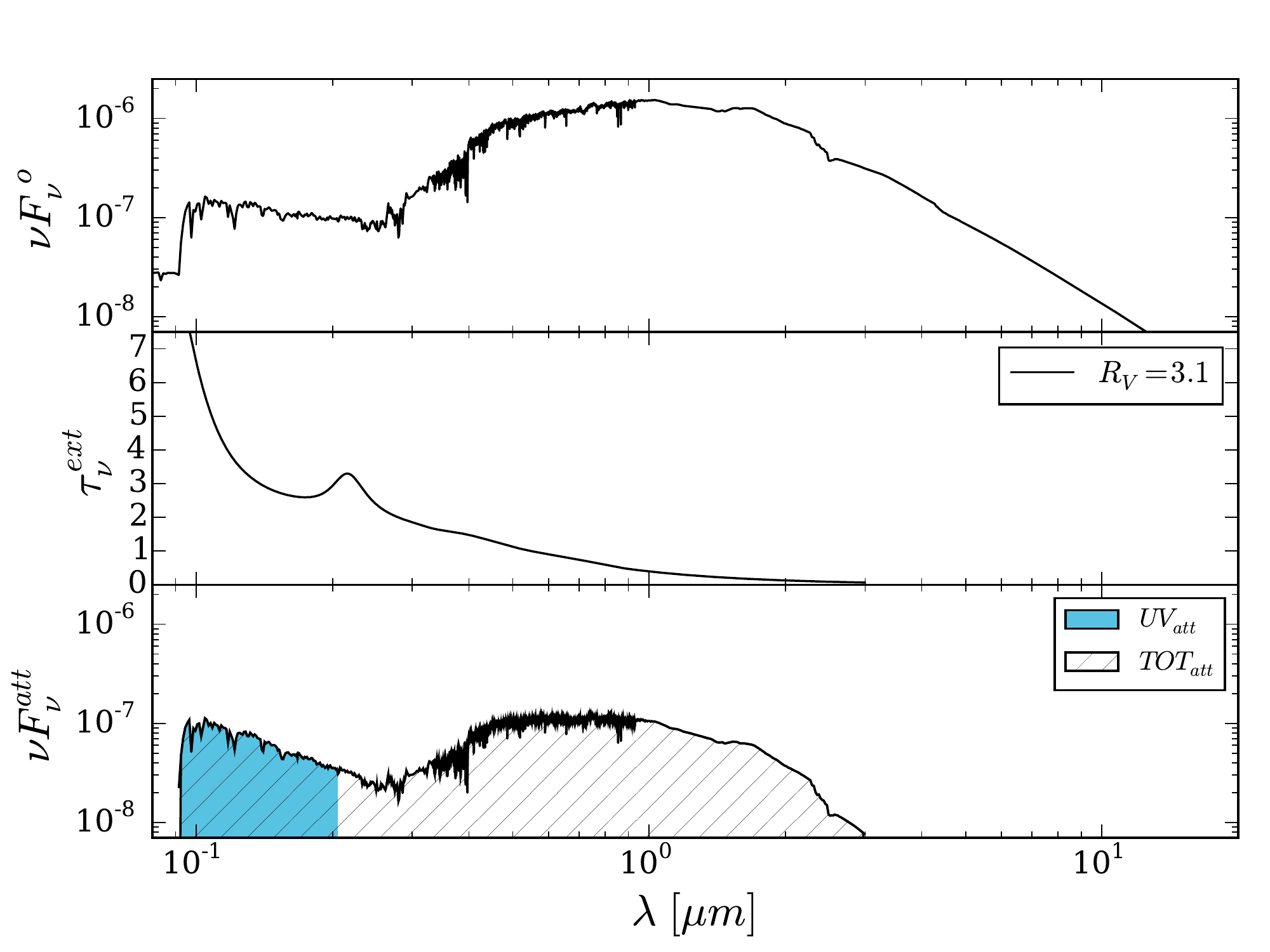}
\caption{Top panel presents the intrinsic stellar spectrum of the entire disk (0.75\degree$\times$3.38\degree; 10$\times$46\,kpc) of M\,31 resulting from the {\tt MAGPHYS} fit to the photometric measurements: GALEX FUV and NUV, SDSS {\it griz}, Spitzer IRAC and MIPS\,24\,\mum, and all Herschel bands. The middle panel shows an example extinction curve \citep[$R_V=3.1$;][]{Cardelli}. The bottom panel presents the attenuated energy from {\tt MAGPHYS} in the 91.2--206.7\,nm ($\rm UV_{att}$), and 91.2--3000\,nm wavelength range. Conceptually, the bottom panel is ${\rm \nu F_{\nu}^{att} = (1-\exp(-\tau_{\nu}))\times \nu F_{\nu}^{\, o}}$, however the attenuation by dust ($\tau_{\nu}$) in {\tt MAGPHYS} is defined differently (see Section~\ref{sec:magphys}).}
\label{fig:sed_entire}
\end{center}
\end{figure}

Compared to inferring stellar radiation properties solely from the dust IR emission, this method gives us the advantage of estimating the hardness of the intrinsic stellar radiation field. 
We use these full-SED best fits and we define two energy bins over which we integrate the attenuated spectrum. The first, between 6\,eV and 13.6\,eV, captures the stellar energy input capable of ejecting 
electrons from the dust grain surface ($\rm UV_{att}$; indicated as the blue area in the bottom panel in Figure~\ref{fig:sed_entire}).
The second, between 0.01\,eV and 13.6\,eV (where the $10^5$\,nm limit ensures capturing the energy at long wavelengths), measures the total stellar energy absorbed and scattered by dust ($\rm TOT_{att}$; 
marked as the hatched area in the bottom panel in Figure~\ref{fig:sed_entire}).
Their ratio ($UV_{att}/TOT_{att}$) represents an estimate of the fraction of stellar energy that can contribute to gas heating  
compared to the total stellar energy contributing to the dust heating.

To estimate the uncertainties in $\rm UV_{att}$ and $\rm TOT_{att}$, we incorporated them into the {\tt MAGPHYS} code.
The parameters, $\rm UV_{att}$ and $\rm TOT_{att}$, are not used explicitly in the {\tt MAGPHYS} code, but by calculating these parameters simultaneously with the model fit, we can build the Bayesian likelihood distributions for them, in the same way as for any other physical parameter in {\tt MAGPHYS}.

\begin{figure}[!ht]
\begin{center}
\includegraphics[width=.5\textwidth]{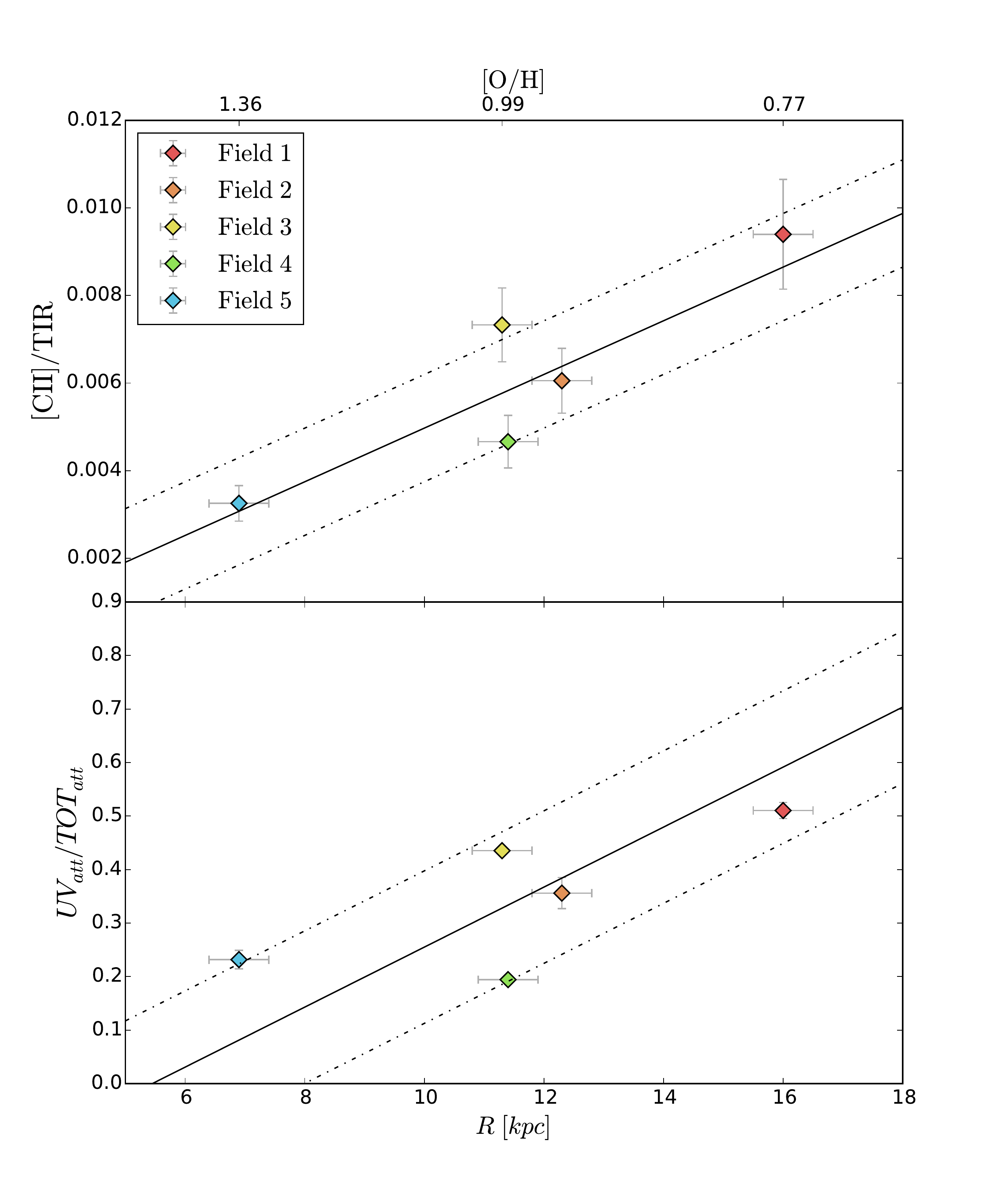}
\caption{Top panel shows the average [\cii]/TIR surface brightness ratio, an often used proxy for gas heating efficiency, against the galactocentric radius (bottom x-axis) and the average gas-phase metallicity (top x-axis; \citet{Zurita2012}'s oxygen abundance as a proxy for metallicity in units relative to solar ${\rm (O/H)/(O/H)_{\odot})}$ for the five SLIM Fields \citep{Kapala2015}. Bottom panel presents $\rm UV_{att}$ over $\rm TOT_{att}$ based on {\tt MAGPHYS} SED fitting versus galactocentric radius. The black solid lines indicate best fits to the data, while dashed-dotted lines show 1-$\sigma$ scatter around best fits. The lines serve merely to guide the reader's eye.}
\label{fig:UVatt_TOTatt_radius}
\end{center}
\end{figure}

We note that {\tt MAGPHYS} is typically applied to integrated galaxy SEDs, and works under the assumption that energy is conserved, allowing for a small variation to account for the effects of inclination.
In this work we use {\tt MAGPHYS} on $\sim$700\,pc scales, which is comparable to pixel-by-pixel ($\gtrsim$659\,pc) SED fitting of 7 nearby galaxies done by \citet{Boquien2012} and \citet{Viaene2014}. 
These scales are large enough that the assumption of energy conservation should be reasonable.
With these caveats in mind, we present the derived $\rm UV_{att}$/$\rm TOT_{att}$ ratio which is an estimate of the relative hardness of the absorbed stellar radiation in Figure~\ref{fig:UVatt_TOTatt_radius}. There appear to be an increasing trend in Figure~\ref{fig:UVatt_TOTatt_radius}, however due to a large scatter and a low number statistics it is not significant (for more details on the relation see Figure~\ref{fig:UVatt_TOTatt}). %, it also depend on local conditions)

%===================================================================

\section{Results}
\label{sec:results}

%-----------------------------------------------------------------

\subsection{Photoelectric heating efficiency estimates}
\label{sec:pe_heating}

Theoretically, the photoelectric heating efficiency is defined as the ratio of the photoelectric heating rate to the rate at which energy in the FUV part of the spectrum is absorbed by dust grains \citep[see e.g.][]{Tielens2008}. 
{\it Ab initio} estimates for the heating efficiency cover the range ${\rm \epsilon_{PE}^{th}} \sim 0.1-5\%$, depending primarily on the local conditions in the gas \citep{Bakes1994, Weingartner2001}. 
Observationally, the most commonly used tracer of the photoelectric heating efficiency is [\cii]/TIR.
However, there is a major discrepancy between these two values as $\epsilon_{\rm PE}^{\rm dust}$ takes into account dust heating by soft optical photons, whereas ${\rm \epsilon_{PE}^{th}}$ considers only FUV photons.

%----

In the following we make a new observationally driven estimate ${\rm \epsilon^{UV}_{PE}}$ from the attenuated UV and [\cii] emission, and compare that to $\epsilon_{\rm PE}^{\rm dust}$ to see if the latter is actually tracing PE efficiency variations.

We now take advantage of our {\tt MAGPHYS} results to improve the estimate of the photoelectric heating efficiency. As discussed above, the traditional [\cii]/TIR estimate suffers from a mismatch between the wavelengths responsible for dust heating and those responsible for PE heating. Instead of using TIR, we match explicitly the wavelengths of photons absorbed by dust to the wavelengths responsible for PE. Specifically, we substitute $\rm UV_{att}$ for TIR, to  define a new approximation of the photoelectric heating efficiency as:
\label{eq:cii_uv}
\begin{equation}
\label{eq:pe_uv}
\epsilon_{\rm PE}^{\rm UV} = \frac{\eta_{[CII]} \Gamma_{gas}(6-13.6\,eV)}{\Gamma_{UVabs}(6-13.6\,eV)} \simeq \frac{[CII]}{UV_{att}}. 
\end{equation}
$\epsilon_{\rm PE}^{\rm UV}$ is closer to the theoretical photoelectric heating efficiency, although we still need to approximate the total gas heating by the [\cii] emission, and $\rm UV_{att}$ includes both scattering and averaging over the full grain size and charge distributions. However, the latter two averages are accounted for in standard calculations of the efficiency as well.

By accounting for only the absorbed FUV photons, the approximation in Equation~\ref{eq:pe_uv} marginalizes out the local stellar photon energy distribution, allowing us to see more clearly the impact of grain properties on the PE efficiency.

%-----------------------------------------------------------------

\subsection{PE efficiency in the SLIM fields}
\label{sec:pe_slim}

\begin{figure}[!ht]
\begin{center}
\includegraphics[width=.5\textwidth]{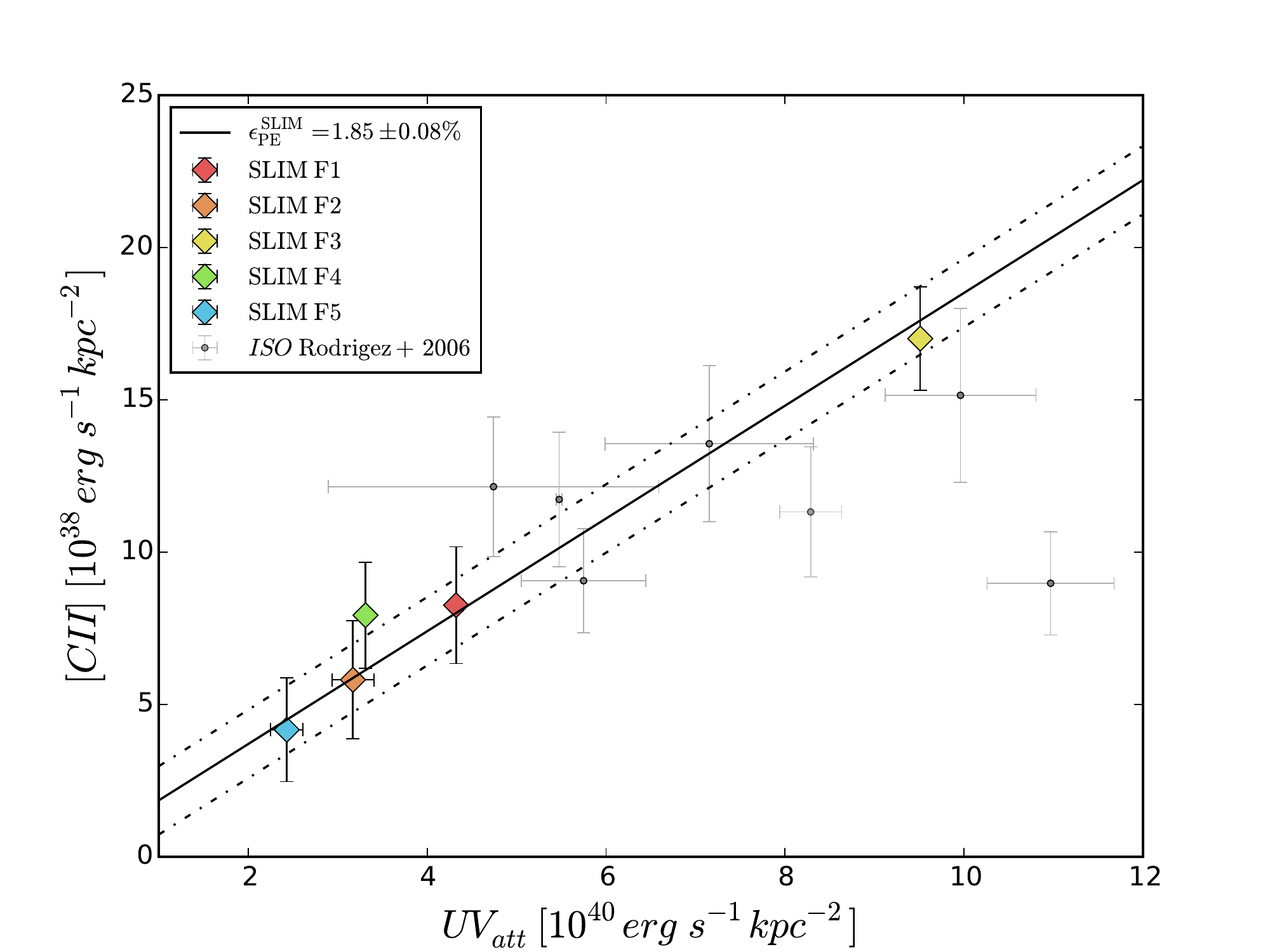}
\caption{Observed [\cii] emission in M\,31 versus predicted attenuated FUV emission, $\rm UV_{att}$, derived from the {\tt MAGPHYS} SED fitting. Based on the SLIM fields (diamonds) we infer a photoelectric heating efficiency $\epsilon_{\rm PE}^{\rm UV} (SLIM)=1.85\pm 0.08\%$. Our best fit is indicated as a solid line, while the dashed-dotted lines indicated the 1-$\sigma$ scatter. To increase the sample size we add averaged {\em ISO} measurements \citep{Rodriguez2006} over SLIM-sized regions (gray points), which are consistent with our inferred fit within 1-$\sigma$, except for one outlier point. }
\label{fig:CII_UVatt}
\end{center}
\end{figure}

Before calculating $\epsilon_{\rm PE}^{\rm UV}$, we first plot the observed [\cii] emission versus the modelled $\rm UV_{att}$ from {\tt MAGPHYS} in the SLIM fields in Figure~\ref{fig:CII_UVatt}. We find a very good correlation.
A linear fit to  [\cii]$\,=\epsilon_{\rm PE}^{\rm UV} \times UV_{att}$, using orthogonal distance regression \citep[ODR from python.scipy package;][]{Brown1990} that accounts for uncertainties along both $x$ and $y$ axis, results in a photoelectric heating efficiency $\epsilon_{\rm PE}^{\rm UV} (SLIM)=1.85\pm 0.08\%$.
We obtain a similar answer, $\epsilon_{\rm PE}^{\rm UV} (SLIM)=1.93\pm 0.24\%$, when we simply average $\epsilon_{\rm PE}^{\rm UV}$ values for the 5 SLIM fields using  Equation~\ref{eq:pe_uv}. 

These values agree well with theoretical predictions. 
We expect to recover efficiencies in the cold neutral medium (CNM) that are close to the theoretical maximum of $\sim 5$\%, but to find much lower values in the WNM ($\sim 1$\%) and in dense PDRs \citep[$\sim 0.3$\%; see e.g.][]{Tielens2008}.
Our result therefore suggests that we are probing emission from a mixture of these three components within each SLIM field.
The resolved [\cii] maps of the SLIM fields confirm that picture, with a few star-forming regions surrounded by more quiescent regions \citep[see Figures 2 and 3 in][]{Kapala2015}.

Given that our result is based on only 5 points, this result is clearly limited by low number statistics. To test the robustness of our PE efficiency, we include archival [\cii] \emph{Infrared Space Observatory} (\emph{ISO}) measurements \citep{Rodriguez2006}. 
Unfortunately, the regions targeted by \citet{Rodriguez2006} lie away from the major axis, on the far side of the M\,31's 10\,kpc ring. Therefore these likely suffer larger projection effects and energy conservation issues compared to the SLIM fields.
We show the additional points as gray circles in Figure~\ref{fig:CII_UVatt}. All but one point agree with our PE efficiency at 1-$\sigma$.  If all points are included in the fit we would derive $\epsilon_{\rm PE}^{\rm UV} (all)=1.55\pm 0.14\%$.

If other mechanisms dominate the heating of the ISM from which the [\cii] emission arises (such as turbulent shocks) then our determined PE efficiency will overestimate the true value. However, \citet{Kapala2015} found only a small contribution of  ionized gas to the total [\cii] emission in the SLIM fields, based on [\nii] 122\,\mum\ emission line observations. 
Similarly, based on diagnostics such as linewidths, X-ray emission, cosmic-ray emission and modelling (Kapala et al. in prep), we find that other mechanisms that can heat the neutral ISM (i.e. cosmic rays, X-rays, shocks/mechanical heating) contribute negligibly to the [\cii] emission in the SLIM fields, therefore we ignore them hereafter.
Given the small ionized gas contribution to [\cii] and
the minor contribution of other heating mechanisms, we expect [\cii] emission in these fields to be generated mostly by photoelectric heating. 
Alternatively, $\epsilon_{\rm PE}^{\rm UV}$ could be an underestimate if [\cii] does not dominate neutral gas cooling.
Based on the relatively small [\oi] contribution to the cooling in the SLIM fields \citep{Kapala2015}, it is very likely that [\cii] is the dominant coolant.

Given that photoelectric heating dominates the [\cii] emission, Figure~\ref{fig:CII_UVatt} demonstrates that a constant photoelectric heating efficiency 
(i.e. $\epsilon_{\rm PE}^{\rm UV}\sim1.85\%$)  can accurately predict  the observed [\cii] emission in all of the SLIM fields. Therefore, the factor of 3 gradient
across the disk in [\cii]/TIR (see  Figure 9 in \citet{Kapala2015})  \emph{can not be} related to the PE efficiency. 
That raises the question: what else drives the [\cii]/TIR gradient in the disk?

%-----------------------------------------------------------------

\subsection{$UV_{att}/TOT_{att}$ correlation with [CII]/TIR}
\label{sec:uv_tot_att_slim}

A major weakness in the [\cii]/TIR ratio as a PE heating efficiency proxy is its dependence on the relative shape of the stellar radiation field. 
To quantify the relative hardness of the dust-absorbed stellar radiation field, we use our {\tt MAGPHYS} output parameters $\rm UV_{att}$ and $\rm TOT_{att}$. The ${\rm UV_{att}/TOT_{att}}$ ratio is an estimate of the fraction of stellar energy that can contribute to gas heating over the total absorbed stellar energy.

\begin{figure}[!ht]
\begin{center}
\includegraphics[width=.5\textwidth]{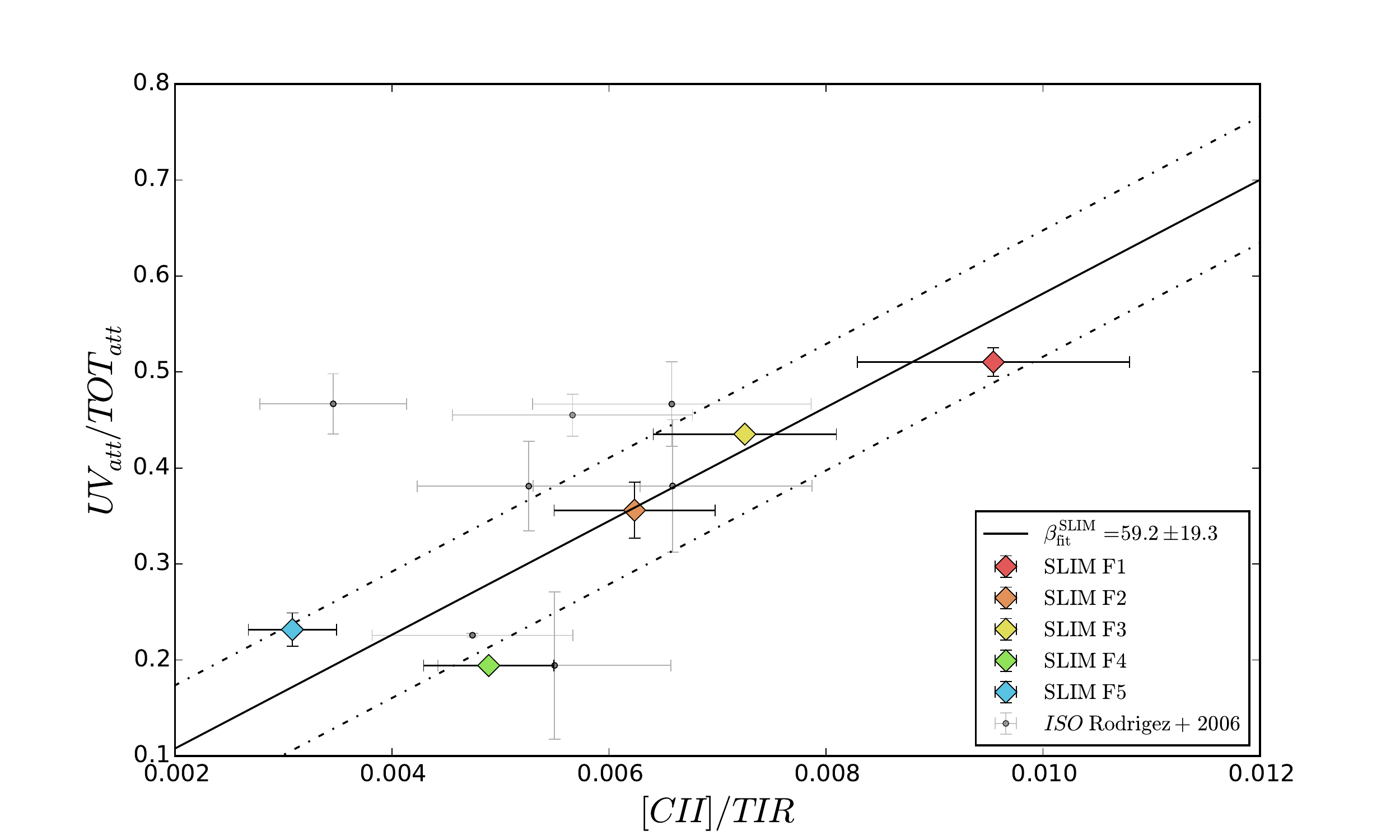}
\caption{Measured [CII]/TIR ratio versus derived ratio of the $\rm UV_{att}$ over $\rm TOT_{att}$ from {\tt MAGPHYS} SED fitting. The symbols and lines indicate the same as in Figure~\ref{fig:CII_UVatt}. This suggests that the relative hardness of the absorbed stellar radiation is a driving factor of the [CII]/TIR ratio.}
\label{fig:UVatt_TOTatt_CIIoverTIR}
\end{center}
\end{figure}

We find a correlation between  [\cii]/TIR and  ${\rm UV_{att}/TOT_{att}}$ in the SLIM fields (Figure~\ref{fig:UVatt_TOTatt_CIIoverTIR}). 
This correlation emphasizes the discrepancy between the theoretical PE efficiency definition and [\cii]/TIR approximation, and supports the argument 
that variations in [\cii]/TIR are primarily driven by variations in the hardness of the dust-absorbed stellar radiation field in the disk of M\,31.

Following Figure~\ref{fig:CII_UVatt}, we populate the ${\rm UV_{att}/TOT_{att}}$ vs [\cii]/TIR plot with averaged [\cii] observations by \citet{Rodriguez2006} over SLIM-sized regions. 
Here, in Figure~\ref{fig:UVatt_TOTatt_CIIoverTIR}, we again see agreement at the 1-$\sigma$ level for all but one of the {\em ISO} points, despite the projection effects and energy conservation problems that likely affect those regions of M\,31.

%===================================================================

\section{Discussion}
\label{sec:disc}

The [\cii]/TIR ratio is the most commonly used observational proxy for PE heating efficiency. This ratio has been found to be ``deficient'' in many galaxies on global and sub-galactic scales suggesting large variations in the PE efficiency \citep[see the discussion section in][for an overview]{Smith2016}. 
Our observation \citep{Kapala2015} that the [\cii]/TIR ratio changes by a factor of $\sim$3 in M\,31, while [\cii] still closely traces SFR suggested that  [\cii]/TIR might not be a good proxy for the PE efficiency. 
The correlation of [\cii]/TIR with ${\rm UV_{att}/TOT_{att}}$ in Figure~\ref{fig:UVatt_TOTatt_CIIoverTIR} indicates that, in M\,31 at least, one of the drivers for [\cii]/TIR variation is the relative shape of the absorbed stellar radiation field.

Studies of other galaxies have attributed [\cii]/TIR variations to other sources, none of which are a plausible cause for M\,31's [\cii]/TIR behavior. 
The variation in the fraction of the total dust in PAHs, i.e. q$_{\rm PAH}$, across M\,31 \citep{Draine2014} peaks in dense ISM structures such as the 10\,kpc ring, therefore q$_{\rm PAH}$ does not correlate with the radially changing [\cii]/TIR. This suggests that destruction of PAHs is minimal and is not impeding PE heating.  Ionized gas contributions leading to overestimates of cooling have been ruled out by [\nii] observations \citep{Kapala2015}. The influence of an AGN on gas ionization or infrared emission is negligible at this distance from the center and given M\,31's lack of a strong AGN \citep{Li2011}.

The other possible driver of the [\cii]/TIR variation is radial gradient in gas-phase metallicity, already considered as a potential explanation in \citet{Kapala2015}. 
Following the arguments proposed by \citet{Israel1996,Israel2011}, a lower metallicity is associated with a lower dust-to-gas ratio. 
A lower dust-to-gas ratio enables the UV to penetrate more deeply into the clouds \citep{Lebouteiller2012}, but also allows a relatively greater number of the soft photons to escape due to the increased mean free path of photons due to the decreased opacity of the gas.
Thus the absorbed radiation field in a volume of gas would become relatively harder, and this could explain the observed increase in the [\cii]/TIR ratio with galactocentric radius in M\,31.
\citet{Israel2011} also describe a counter-mechanism, where lower metallicity means also lower PAH abundances, and therefore decreases gas heating efficiency.     
Although the first mechanism goes in the right direction to explain our observations of [\cii]/TIR increasing with decreasing metallicity, \citet{Kapala2015} could not explain the amplitude of the observed variations. 
The difference in the metallicities of the different SLIM fields is relatively small, ranging from $\sim 0.77$~Z$_{\odot}$ to $\sim 1.36$~Z$_{\odot}$.
We note that the metallicities in the Magellanic Clouds are significantly lower ($\sim$\,0.5 and $\sim$\,0.25\,Z$_{\odot}$) than in M\,31, therefore  the impact of metallicity on the ISM is likely to be stronger in the MCs than M\,31.
Furthermore, considering the correlation with the absorbed stellar radiation hardness (Section~\ref{sec:uv_tot_att_slim}) and SFRs \citep{Smith2016}, the metallicity seems to be a secondary factor that impacts [\cii]/TIR or [\cii]/$\rm UV_{att}$ in massive, star forming galaxies. 
Nevertheless, since the $\rm UV_{att}$ is defined as an integral over a narrower wavelength range than TIR, ${\rm \epsilon^{UV}_{PE}}$ should be less affected as a PE heating approximation than ${\rm \epsilon^{dust}_{PE}}$ by metallicity changes.

In Section~\ref{sec:uv_tot_att_slim} we presented an  explanation of the [\cii]/TIR variation within M\,31 as driven by the ${\rm UV_{att}/TOT_{att}}$. However the relative change in absorbed stellar hardness is controlled by more than one parameter, mainly by the star formation history (SFH) and dust opacity/extinction.
While we cannot quantify the contributions of SFH and dust opacity to the ${\rm UV_{att}/TOT_{att}}$ changes across M\,31, the correlations we find indicate that changes in the star formation history seems to  dominate (for details see Appendix~\ref{sec:mag_corr} where we explore  correlations of other parameters with the ${\rm UV_{att}/TOT_{att}}$ ratio).

It is reassuring that our findings are consistent with \citet{Kapala2015} using an independent method, a spatial comparison between various SFR tracers: H$\alpha$, [\cii], 24\,\mum\ and TIR. \citet{Kapala2015} looked at how compact/extended the emission of these tracers is around star-forming regions, and found that H$\alpha$ is the most compact, [\cii] emission is typically more extended than H$\alpha$ but less than 24\,\mum\ and TIR. They suggested that TIR might be affected by heating arising from older stellar populations. Therefore, \citet{Kapala2015} and the current paper agree with each other in the context of TIR being affected by softer radiation coming from older stars.

%........................

There are no independent observational measurements in the literature of the photoelectric efficiency that do not simply rely on [\cii]/TIR, that we know of.
We can try to calculate the PE efficiency for the Milky Way (MW) based on \citet{Pineda2013}, who estimate that roughly half of the observed [\cii] emission in the Milky Way (MW) comes from dense PDRs, and the remainder comes from cold atomic or CO-dark molecular gas. We emphasize that these estimated fractions rely on detections of velocity resolved \emph{Herschel} HIFI spectra that are biased towards bright, SF regions in the MW that tend to be PDR dominated. A similar bias occurs in selection of our SLIM fields in M\,31.
If we take the values for the PE efficiency in dense PDRs and in the CNM from \citet{Tielens2005} of 0.3\% and 3\% respectively, and the \citet{Pineda2013} estimate of the fractional [\cii] contributions from those phases, this results in a PE efficiency of  $\epsilon_{\rm PE} \sim 0.5 \times 0.03 + 0.5 \times 0.003 = 0.0165 = 1.65\%$. This value is in good agreement with our determinations of the efficiency in M\,31.  While this is only a crude estimate for the MW value, the agreement with M\,31 is encouraging considering the similarity between these two galaxies. 

However, the dust SED modeling from \citet{Draine2014} in M\,31 estimates that the SLIM fields have maximal fraction of the total dust luminosity that is radiated by dust grains in PDR regions, f$_{\rm PDR}$\footnote{f$_{\rm PDR}$ is a model parameter, and it can be derived from eq. 29 in \citet{DraineLi2007}, using U$_{\rm min}$ and $\gamma$ maps from \citet{Draine2014}, where U$_{\rm max}$ is fixed at $10^7$. These maps can be downloaded from  www.astro.princeton.edu/\~{}draine/m31dust/m31dust.html}, values in the range of 5-26\%, significantly lower than the MW estimates. The PDR fractions agree better with \citet{Croxall2012} who estimated the fraction of the [\cii] emission arising from PDRs to be less than 25\% for two nearby galaxies (NGC 1097 and NGC 4559). If we use this lower estimate of 25\% PDR fraction in our determination of the MW PE efficiency, we get $\epsilon_{\rm PE} \sim 0.75 \times 0.03 + 0.25 \times 0.003 = 0.02325 \sim 2.33\%$. This value is above our estimate for $\epsilon_{\rm PE}$ in M\,31, but still reasonably close. Based on this, our measured PE for M\,31 is consistent with theoretical expectations over a wide range of potential PDR/CNM [\cii] fractions.

\begin{figure}[!ht]
\begin{center}
\includegraphics[width=.5\textwidth]{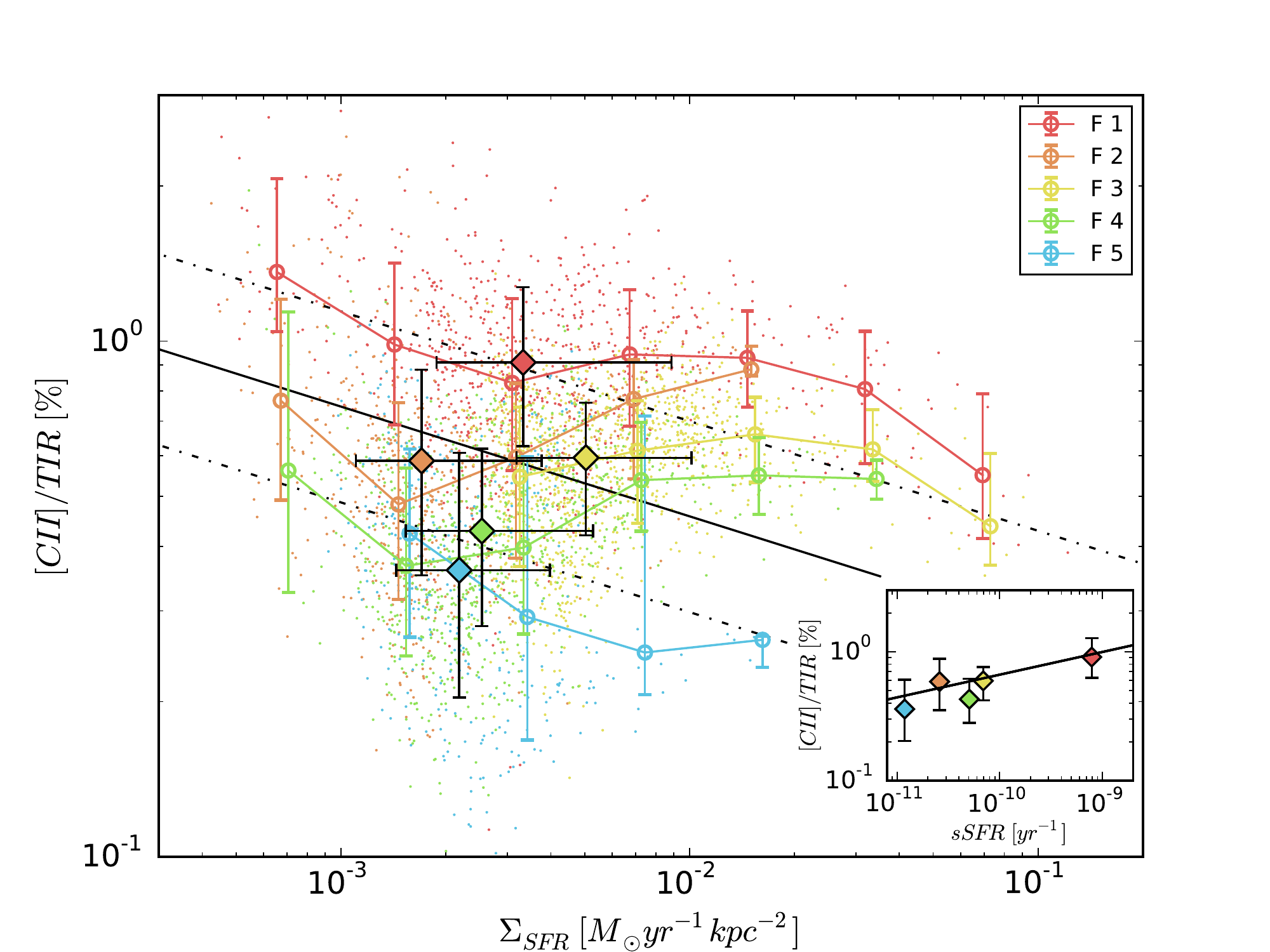}
\caption{Measured [CII]/TIR ratio versus $\Sigma_{\rm SFR}$. Point symbols show all of the significantly detected points in SLIM fields. Circles indicate median [CII]/TIR, while error bars indicate 16$^{\rm th}$ and 84$^{\rm th}$ percentiles within evenly spaced $\Sigma_{\rm SFR}$ bins in a log scale. The median values in the fields are marked as diamonds, error bars present 16$^{th}$ and 84$^{th}$ percentiles. Black solid and dashed-dotted lines indicate a power-law fit and fit uncertainty, respectively,  to the binned KINGFISH regions, GOALS galaxies, and high redshift sources \citep[equation 1 and figure 3 in][]{Smith2016}. {\it Bottom right corner:} Diamonds show [CII]/TIR relation with sSFR (SFR per unit stellar mass) for SLIM fields. The black fit is derived from the sSFR--${\rm UV_{att}/TOT_{att}}$ and ${\rm UV_{att}/TOT_{att}}$--[CII]/TIR relations. The advantage is that the former relation is more statistically significant (29 measurements across galaxy comparing to 5 points).}
\label{fig:CIIoverTIR_SFR}
\end{center}
\end{figure}

However, most studies use [\cii]/TIR to estimate the PE efficiency. The recent work of \citet{Smith2016} examining the [\cii]/TIR ratio in nearby galaxies, has a sample (KINGFISH galaxies) close in resolution and stellar mass to M\,31. They find a  trend of decreasing [\cii]/TIR with increasing $\Sigma_{\rm SFR}$ that becomes strong when measurements from other studies are included.
In Figure~\ref{fig:CIIoverTIR_SFR}, we directly compare the M\,31 [\cii]/TIR data (available only in the SLIM fields) and corresponding $\Sigma_{\rm SFR}$ \citep[derived from H$\alpha +$\,24\mum\ using the][formula]{Calzetti2007} with the \citet{Smith2016} results. 
We find a comparable, large scatter in [\cii]/TIR relation with $\Sigma_{\rm SFR}$ derived from H$\alpha+24$\,\mum\ \citep[][formula]{Calzetti2007}  at  $\sim$\,50\,pc scales (figure~\ref{fig:CIIoverTIR_SFR}) as \citet{Smith2016} at $\sim$\,sub-kpc scales \citep[within 54 nearby galaxies covering a wide range of galaxy properties and local ISM environment; figure~3,][]{Smith2016}.
We see a similar, albeit weaker, declining trend of [\cii]/TIR with $\Sigma_{\rm SFR}$ as \citet{Smith2016} on smaller scales (evenly binned 50\,pc measurements in $\Sigma_{\rm SFR}$ log space; see fig.~\ref{fig:CIIoverTIR_SFR}) than their sub-kpc scales.
When integrated over the SLIM fields, the [\cii]/TIR values (diamonds) show no trend with $\Sigma_{\rm SFR}$, but are consistent with the global trend and scatter found by \citet{Smith2016} (black lines on main plot in fig.~\ref{fig:CIIoverTIR_SFR}).
This lack of a trend may be due to the small dynamic range in $\Sigma_{\rm SFR}$ probed by our SLIM fields, as compared to the 6 orders of magnitude seen in \citet{Smith2016}.

In a similar study of the LMC, \citet{Rubin2009} find [\cii]/TIR roughly constant over $\sim$\,2.5 order of magnitude change of  $\Sigma_{\rm H \alpha}$. However, there are pixels in figure~4 of \citet{Rubin2009} that do appear to follow the same declining trend of \citet{Smith2016}, but were dismissed by the authors as likely affected by the noise in the [\cii] map. 

One obvious conclusion to draw from these comparisons is that the scales and placement of the apertures over which the [\cii]/TIR and $\Sigma_{\rm SFR}$ are measured have a significant impact on the results.

In contrast to the trend with $\Sigma_{\rm SFR}$, given the demonstrated trends of ${\rm UV_{att}/TOT_{att}}$--[\cii]/TIR (fig.~\ref{fig:UVatt_TOTatt_CIIoverTIR}) and sSFR--${\rm UV_{att}/TOT_{att}}$ (fig.~\ref{fig:mag_UV_TOT_1} in appendix~\ref{sec:mag_corr}) relations, we infer that the [\cii]/TIR ratio increases with the specific SFR (sSFR; subplot in the bottom right corner in figure~\ref{fig:CIIoverTIR_SFR}). 
Given that the [\cii]/TIR vs $\Sigma_{\rm SFR}$ relation is in agreement with \citet{Smith2016}, the inverse trend of [\cii]/TIR with sSFR in M\,31 must arise because of the impact of $\Sigma_{\rm M_{*}}$ on the dust heating. As the TIR is affected by the diffuse radiation field arising from all stellar populations, increasing $\Sigma_{\rm M_{*}}$ (and thus decreasing sSFR) must act to increase the TIR and thus decrease [\cii]/TIR, creating the trend seen in figure~\ref{fig:CIIoverTIR_SFR}. Whether this trend still exists in the larger sample of \citet{Smith2016} is yet to be seen.

The \citet{Smith2016} trend of the [\cii]/TIR declining with $\Sigma_{\rm SFR}$ agrees best with the PDR models of \citet{Tielens1985} and \citet{Wolfire1990}.
These models find that the environmental conditions found in extreme star forming regions, such as the increased radiation fields and gas temperatures and densities, all act to relatively decrease the [\cii] emission. This occurs through;
(1) the [\cii] emission line no longer being the dominant coolant, in which case we might test by observing the relative brightness of the [\oi] 63\,\mum\ emission line \citep[this is theoretically predicted][ but has not yet been observed]{Tielens1985}, and 
(2) increasing the average grain charge, therefore increasing the work potential and decreasing the true PE efficiency, which we can test for example by observing PAH features in the MIR \citep{Okada2013}.
It would be interesting to test whether some/all of the trend  goes away if we use our $\epsilon_{\rm PE}^{\rm UV}$ instead of $\epsilon_{\rm PE}^{\rm dust}$ for the larger data sets like \citet{Smith2016}.
If the trend remains, it means that we either observe a real PE efficiency variations or other heating mechanisms or cooling lines take over, where the latter can be tested.

When parameters such as [\cii]/TIR are integrated over a region, any small scale variations, such as the [\cii] and TIR arising from distinct regions, will be smoothed over. It is only through simulations and comparing nearby objects, such as the MW and M\,31, to the distant objects examined in \citet{Smith2016}, can we begin to account for these effects.

%===================================================================

\section{Conclusions}
\label{sec:concl}

The main conclusion of this paper is that in M\,31 the [\cii]/TIR variation is not tracing changes in the photoelectric heating efficiency.
Worth noting is that the photoelectric heating efficiency proxy [\cii]/TIR  is defined significantly differently from the theoretical definition. 
In particular, in this work we have demonstrated that the [\cii]/TIR ratio ($\epsilon_{\rm PE}^{\rm dust}$)  is an imperfect approximation for the photoelectric heating efficiency in M\,31.
We define  a new method  to use a SED fitting technique to derive the energy absorbed by dust in the PE heating wavelength range ($\rm UV_{att}$), that together with the [\cii] emission allows us to determine  a more direct estimate of the photoelectric efficiency:
\begin{equation}
{\rm \epsilon_{\rm PE}^{\rm UV} = \frac{[CII]}{UV_{att}} } \nonumber
\end{equation}

Using [\cii] imaging from the Survey of Lines in M\,31 and a library of multiwavelength observations from UV through IR, we have demonstrated that it is possible to predict the [\cii] emission based on the estimated fraction of the stellar energy that 
contributes to gas heating (${\rm \epsilon_{PE}^{UV}\times UV_{att}}$) with a constant photoelectric heating efficiency ($\epsilon_{\rm PE}^{\rm UV}=1.85\pm0.08\%$). 
A constant photoelectric heating efficiency is also supported by fact that [\cii] works as a SFR tracer in each SLIM field despite the 
factor of 3 gradient in the [\cii]/TIR found between these fields \citep{Kapala2015}.

We found that the attenuated UV energy relative to the total attenuated energy (${\rm UV_{att}/TOT_{att}}$) correlates well with the variation in the [\cii]/TIR ratio suggesting 
that it is the hardness of the dust-absorbed stellar radiation field
that is driving the variation in the [\cii]/TIR ratio across this disk, not PE efficiency changes.

The question remains whether the use of attenuated UV energy can correct the ``[\cii]-deficit'' observed in many galaxies, or whether there is still some remaining factors (such as true variations in $\epsilon_{\rm PE}$) which cause this observed deficit. Larger samples with [\cii] observations and ancillary multiwavelength data, such as the KINGFISH  or GOALS samples, may be able answer whether ${\rm \epsilon^{UV}_{PE}}$ is a true constant or does vary across galaxies.

\vspace{1cm}

%===================================================================

%===================================================================

\section*{Appendix}
\appsection

%-----------------------------------------------------------------

\section{{\tt MAGPHYS} SED modeling across M\,31}
\label{sec:sed_fits_disk}

\begin{figure*}[!ht]
\begin{center}
\hspace{1.2cm}
\includegraphics[width=.9\textwidth]{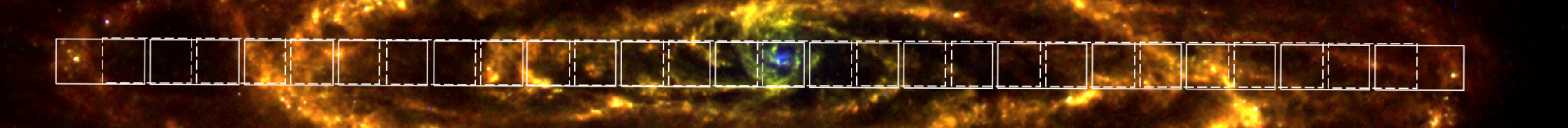}
\includegraphics[width=1.\textwidth]{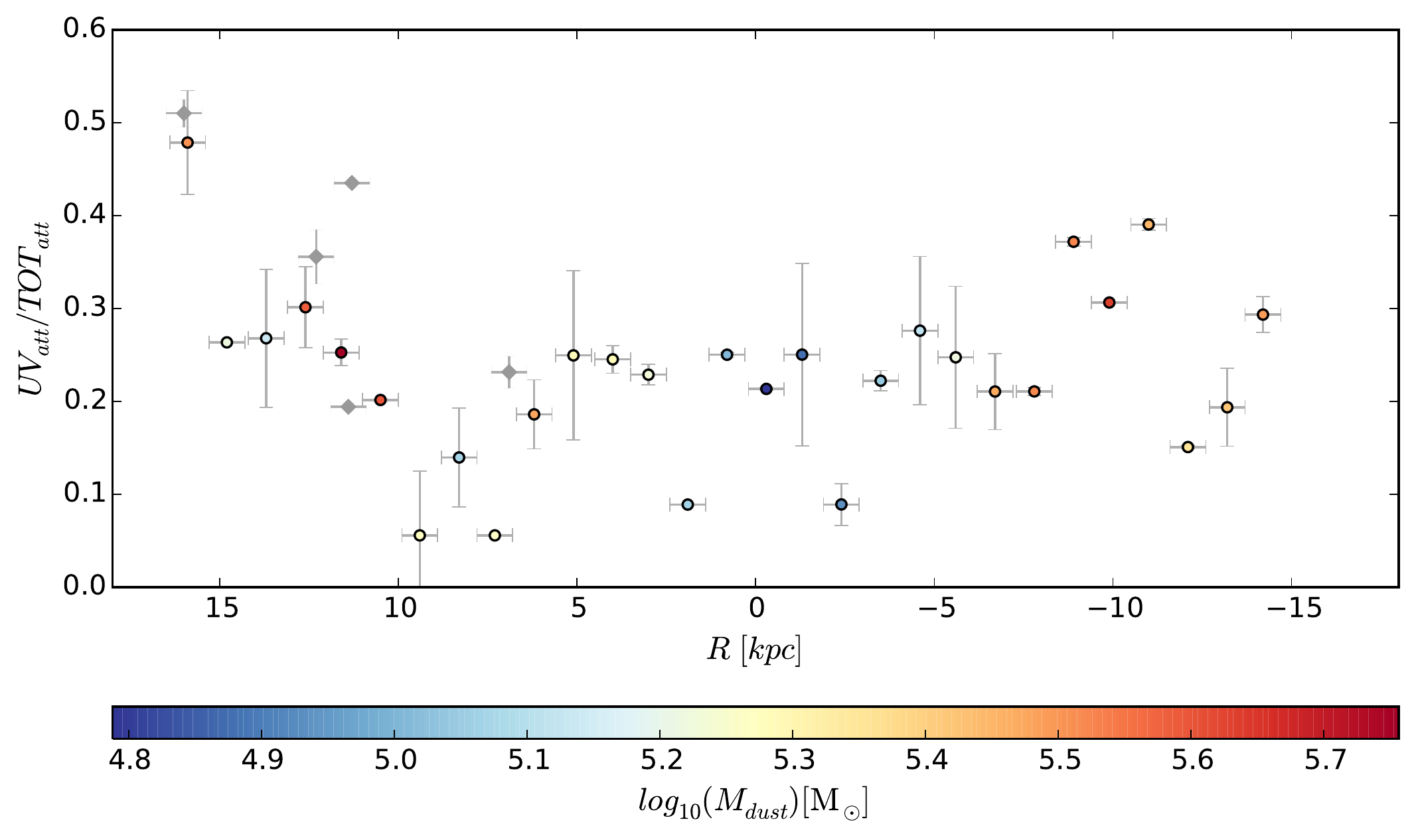}
\caption[${\rm UV_{att}/TOT_{att}}$ versus galactocentric radius]{${\rm UV_{att}/TOT_{att}}$ versus galactocentric radius (negative values indicate SW direction). Circle points are color-coded with dust mass surface density \citep[$M_{dust}$ extracted from][maps]{Draine2014}. Diamond points indicate SLIM fields. If we consider only $\sim$\,6--16\,kpc, there is an increasing trend. However, when all ${\rm UV_{att}/TOT_{att}}$ values on both sides of the galaxy are considered, they are scattered around $0.22 \pm 0.09$.  }
\label{fig:UVatt_TOTatt}
\end{center}
\end{figure*}

We use {\tt MAGPHYS} SED modeling across the galaxy along the major axis in $1 \times 2$\,kpc apertures (dashed and solid boxes in Figure~\ref{fig:m31_1kpc_ap}) to derive
the overall trend of ${\rm UV_{att}/TOT_{att}}$ with galactocentric radius. We present the results  in Figure~\ref{fig:UVatt_TOTatt}. The top panel showing a cut-out of the \emph{Herschel} image of M\,31, and the color-coded points with dust mass
surface density \citep[extracted from][maps]{Draine2014} help to orientate each data point to a position within the galaxy and indicate the structure of the galaxy (e.g., spiral arms). 

There is a hint of the overall radial trend in ${\rm UV_{att}/TOT_{att}}$ on the NE (positive radius) side, however the scatter is considerable.
There is a tentative systematic excess, a relatively higher ${\rm UV_{att}/TOT_{att}}$ in the arm than in the interarm regions, with the exception of the central regions dominated by the massive bulge.

\begin{figure*}[!ht]
\begin{center}
\includegraphics[width=1.\textwidth]{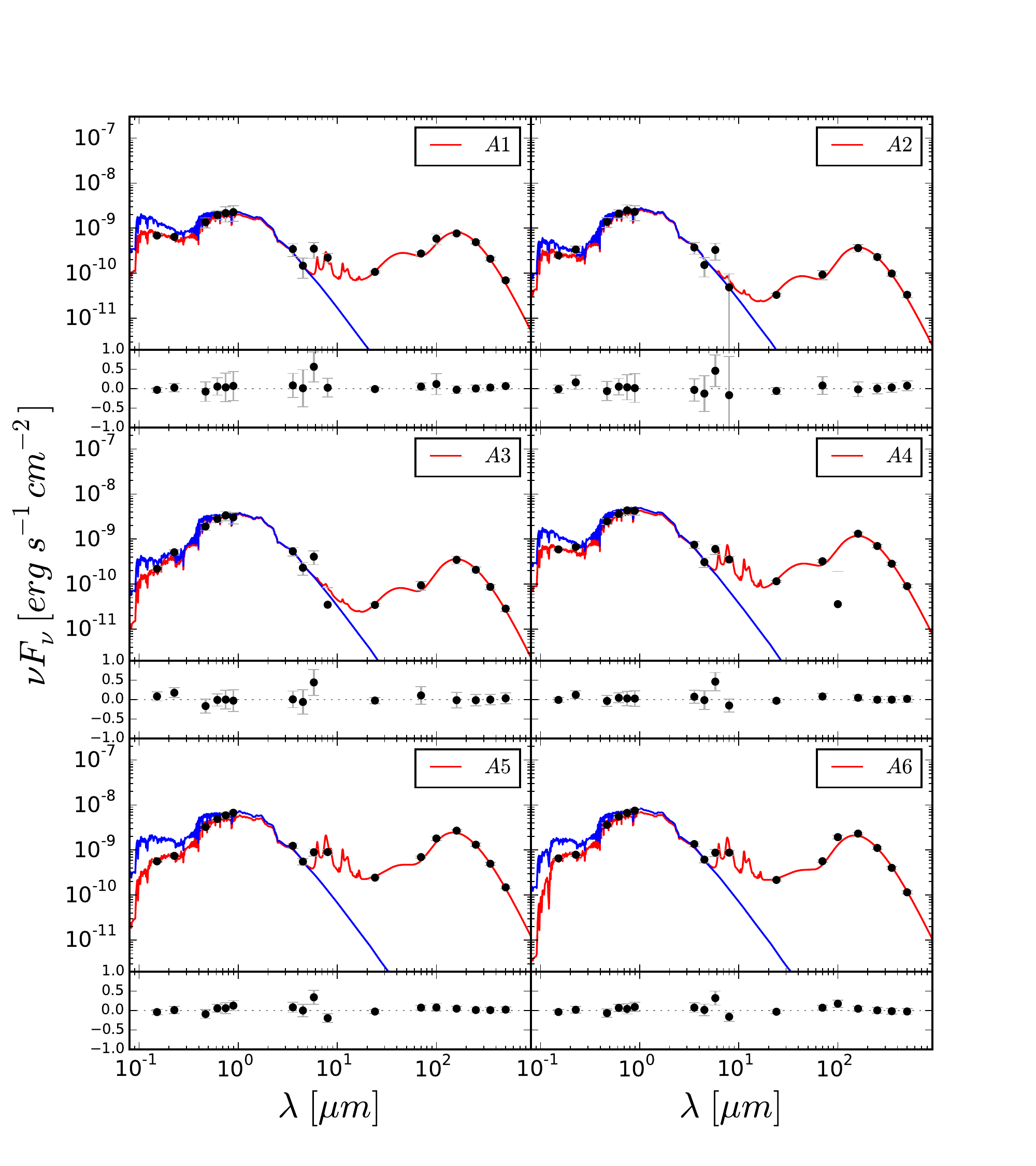}
\caption[The SED fits A1--A6]{The SED across M31 along the major axis, measured within 1\,kpc (4.405\arcmin) apertures A1--A6. The fluxes (GALEX FUV and NUV, SDSS u, g, r, i and z, Spitzer IRAC and MIPS24, and the five Herschel bands excluding SPIRE 500). The red curve is the best fit {\tt MAGPHYS} spectrum to the data (stellar and dust emission), while the blue curve is the inferred unreddened stellar spectrum. Note that A3 is an extremely faint region, and {\tt MAGPHYS} fitting fails there. }
\label{fig:SED0}
\end{center}
\end{figure*}

\begin{figure*}[!ht]
\begin{center}
\includegraphics[width=1.\textwidth]{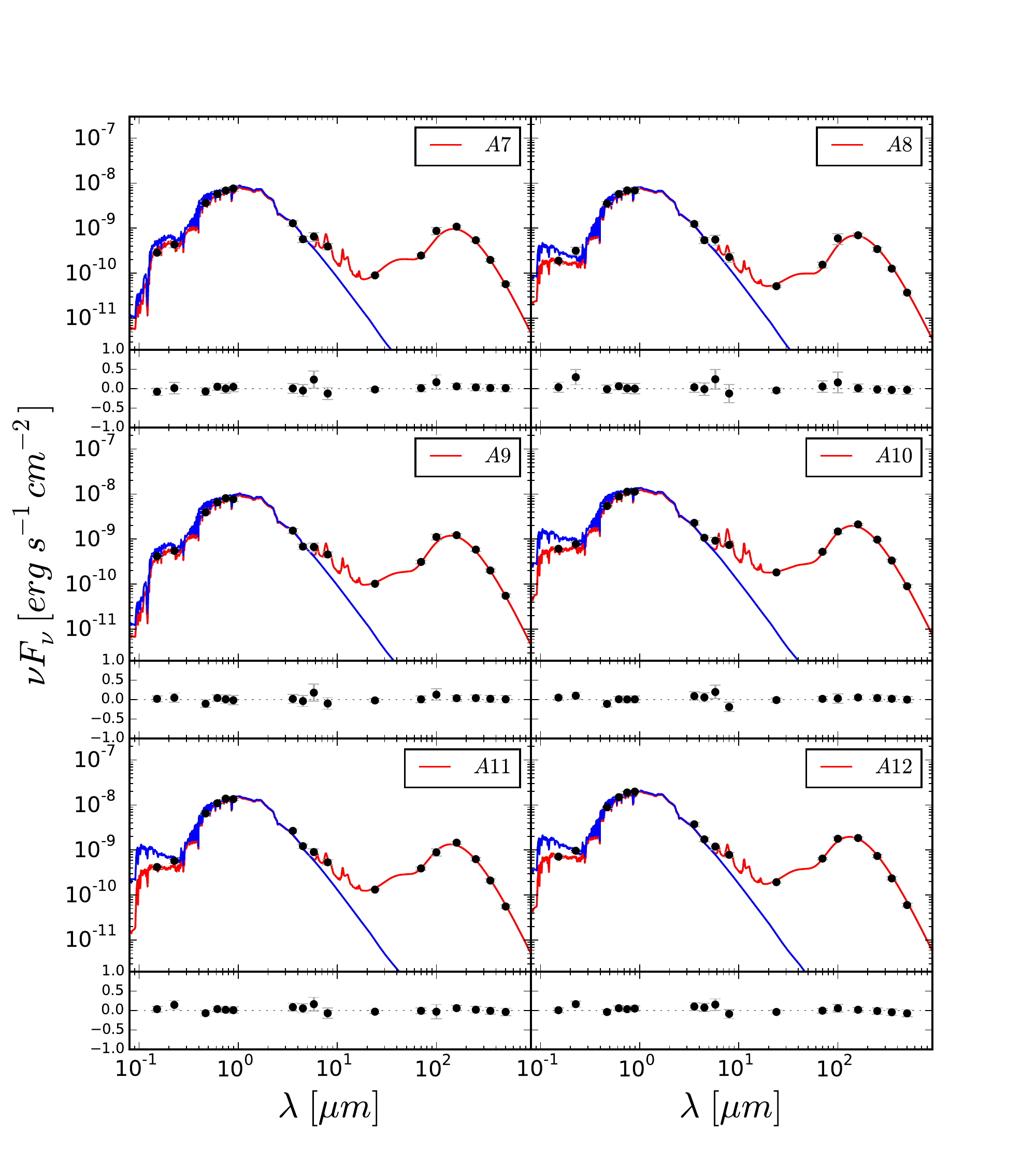}
\caption[The SED fits A7--A12]{As in Figure~\ref{fig:SED0}, but for different apertures A7--A12 as labelled.}
\label{fig:SED1}
\end{center}
\end{figure*}

\begin{figure*}[!ht]
\begin{center}
\includegraphics[width=1.\textwidth]{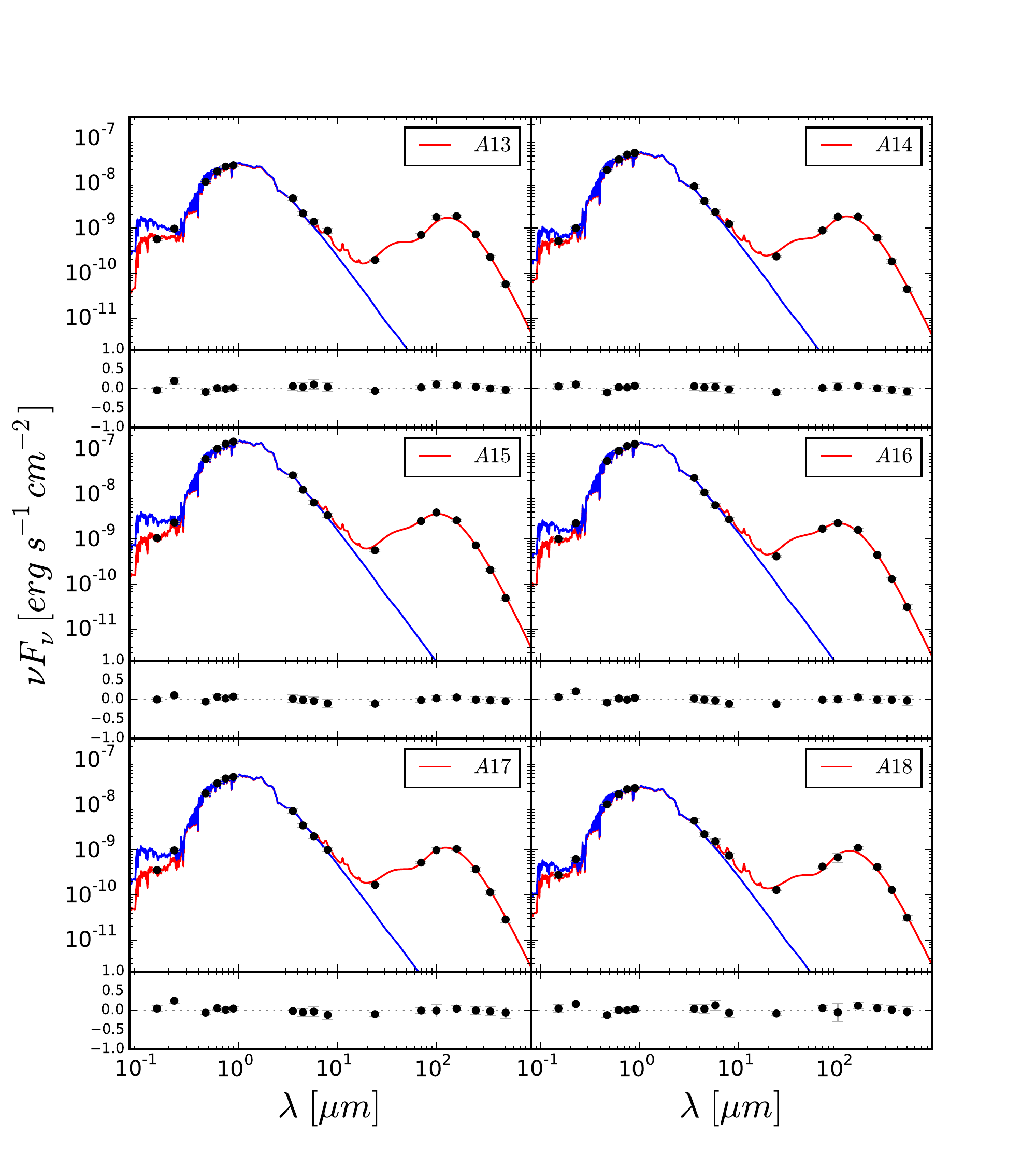}
\caption[The SED fits A13--A18]{As in Figure~\ref{fig:SED0}, but for different apertures A13--A18 as labelled.}
\label{fig:SED2}
\end{center}
\end{figure*}

\begin{figure*}[!ht]
\begin{center}
\includegraphics[width=1.\textwidth]{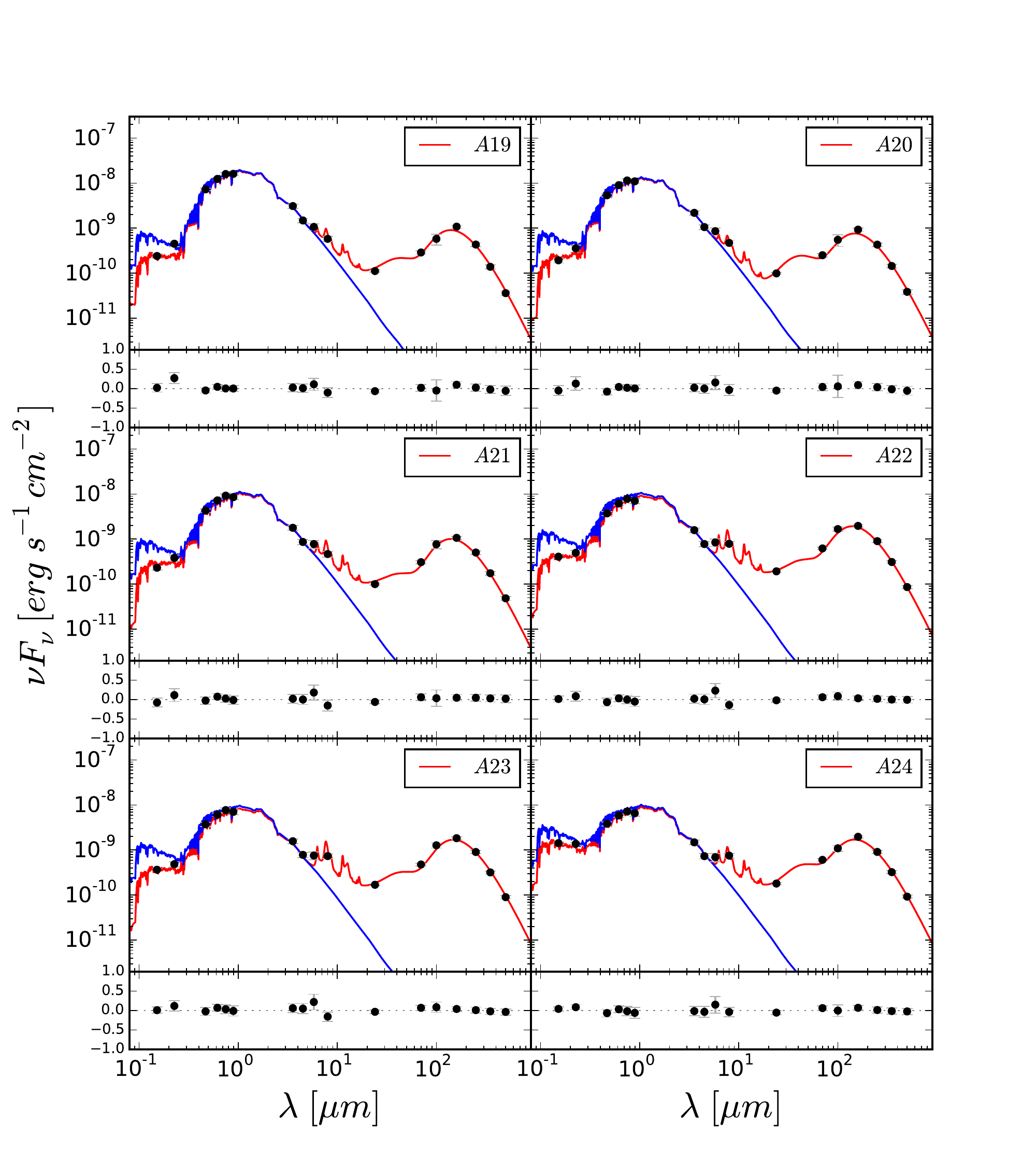}
\caption[The SED fits A19--A24]{As in Figure~\ref{fig:SED0}, but for different apertures A19--A24 as labelled.}
\label{fig:SED3}
\end{center}
\end{figure*}

\begin{figure*}[!ht]
\begin{center}
\includegraphics[width=1.\textwidth]{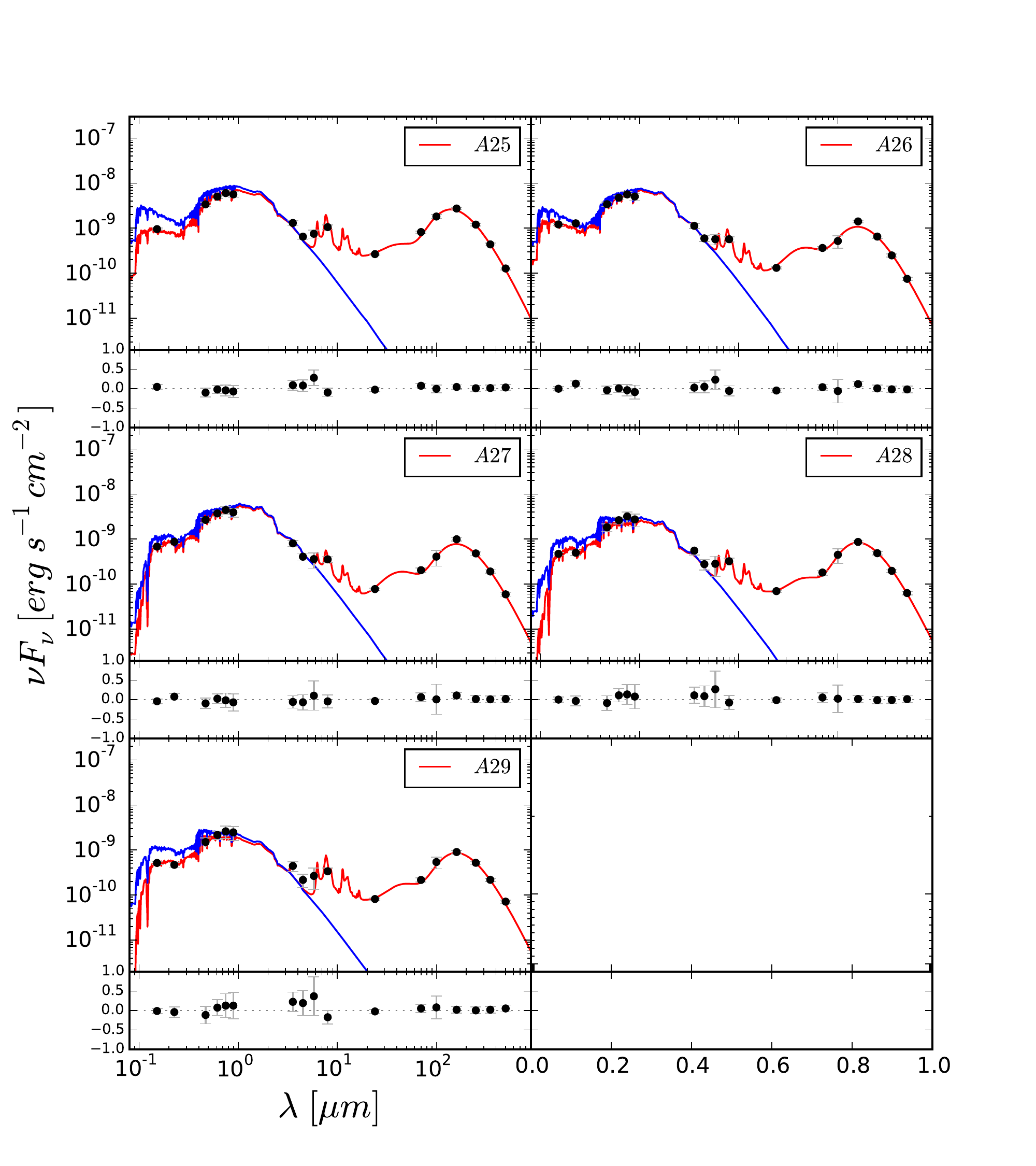}
\caption[The SED fits A25--A30]{As in Figure~\ref{fig:SED0}, but for different apertures A25--A30 as labelled. Note that A27 is an extremely faint region, and {\tt MAGPHYS} fitting fails there.}
\label{fig:SED4}
\end{center}
\end{figure*}

We present the corresponding SED fits for all regions
in Figures~\ref{fig:SED0}--\ref{fig:SED4}. In future work, we plan to use the ${\rm \epsilon_{\rm PE}^{\rm UV}\times UV_{att}}$ to predict [\cii] emission in the disk of M\,31.

%-----------------------------------------------------------------

%===================================================================
%===================================================================

\begin{figure*}[!ht]
\begin{center}
\includegraphics[width=1.\textwidth]{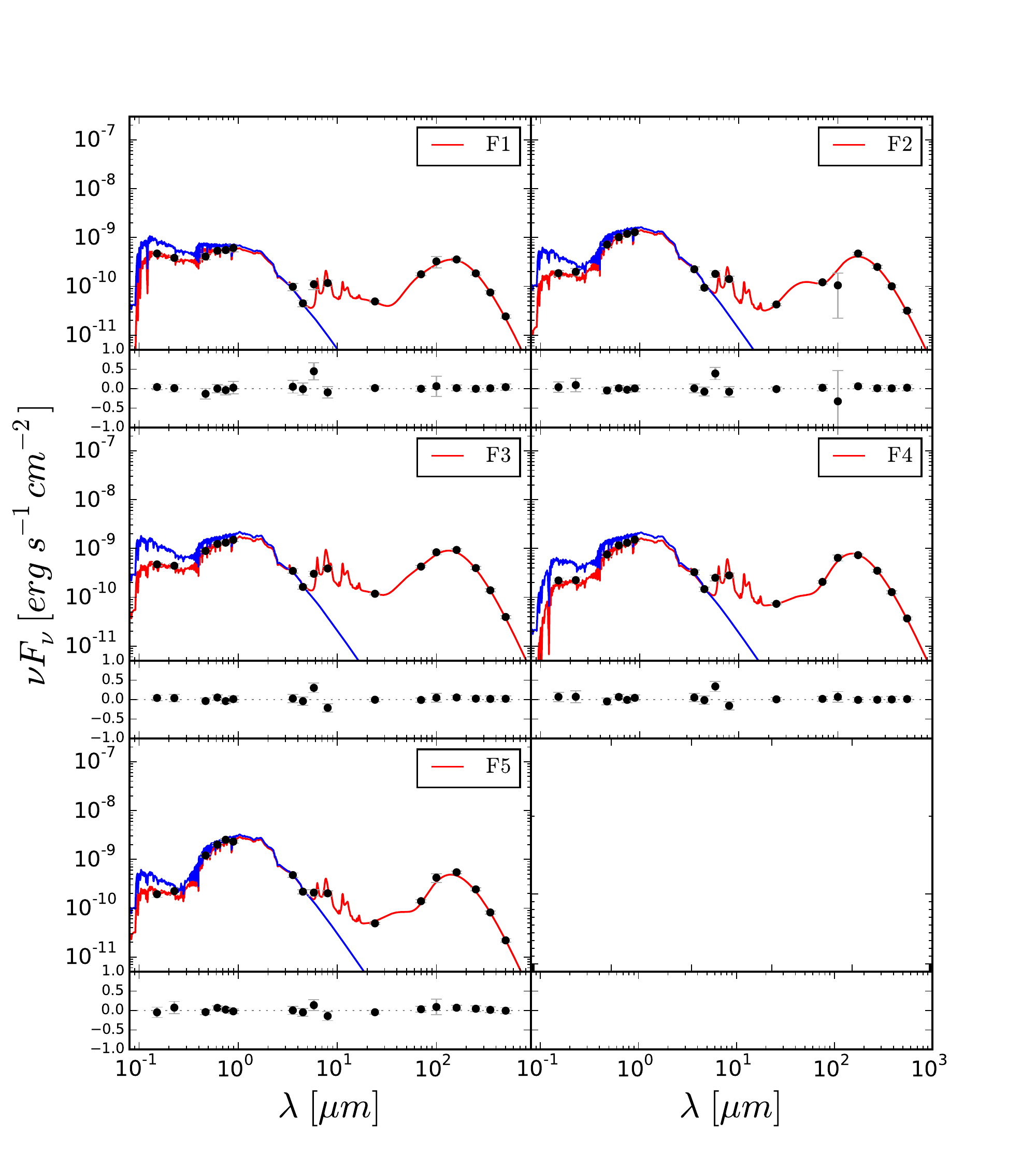}
\caption{{\tt MAGPHYS} SED fits of the SLIM fields F1--F5.}
\label{fig:slim_seds}
\end{center}
\end{figure*}

%===================================================================
%===================================================================

\section{Correlations of {\tt MAGPHYS} parameters}
\label{sec:mag_corr}

We explore correlations between ${\rm UV_{att}/TOT_{att}}$ and various physical and {\tt MAGPHYS} output parameters. 
We find the best anti-correlation with $f_{\mu}$ (the fraction of the total infrared luminosity contributed by the dust in the ambient, diffuse ISM) with a Pearson correlation coefficient $r=-0.86$, giving the probability of an uncorrelated data $p=2.5\times10^{-9}$. This anti-correlation indicates that an increasingly harder radiation is absorbed in the vicinity of star forming regions, or at least where they dominate the IR emission.
${\rm f_{\mu}}$ depends itself on the total effective V-band dust optical depth, ${\rm \tau_V}$ ($r=0.61$, $p=4.3\times10^{-4}$), the fraction, $\mu$, of this contributed by dust in the ambient ISM ($r=-0.38$, $p=0.04$), and  the relative proportion of young and old stars (which we approximate here by sSFR; $r=-0.67$, $p=7.2\times10^{-5}$). 
The second strongest correlation of the ${\rm UV_{att}/TOT_{att}}$ is with sSFR ($r=0.81$, $p=1.1\times10^{-7}$), while with $\mu$ ($r=0.52$, $p=3.2\times10^{-3}$) and $\tau_V$ ($r=-0.34$, $p=0.07$) are much weaker. 
To summarize, while we cannot quantify the contributions of SFH and dust opacity to the ${\rm UV_{att}/TOT_{att}}$ changes, the correlations we find indicate that changes in the star formation history seem to dominate.

\begin{figure*}[!ht]
\begin{center}
\includegraphics[width=1.\textwidth]{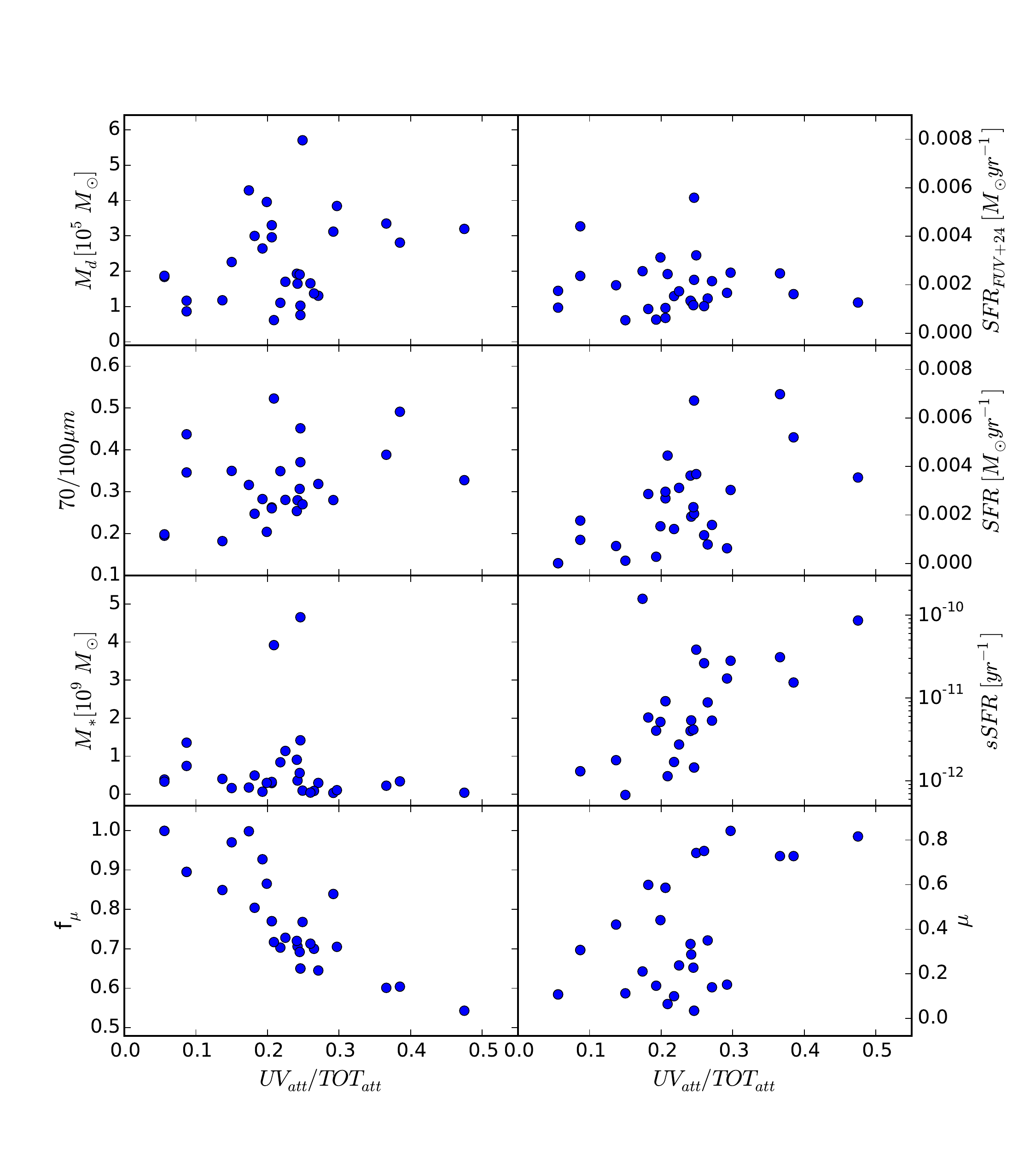}
\caption{Various parameters plotted against $UV_{att}/TOT_{att}$. R, $SFR_{FUV+24}$ and $70/100\mu m$ (dust color) are physical parameters, $M_d$ is the dust mass inferred by \citet{Draine2014}, and the rest are {\tt MAGPHYS} output parameters: SFR, M$_*$, sSFR, $f_{\mu}$ (the fraction of the total infrared luminosity contributed by the dust in the ambient/diffuse ISM), $\mu$ (the fraction of the total effective V-band absorption optical depth
of the dust contributed by dust in the ambient ISM). }
\label{fig:mag_UV_TOT_1}
\end{center}
\end{figure*}

\begin{figure*}[!ht]
\begin{center}
\includegraphics[width=1.\textwidth]{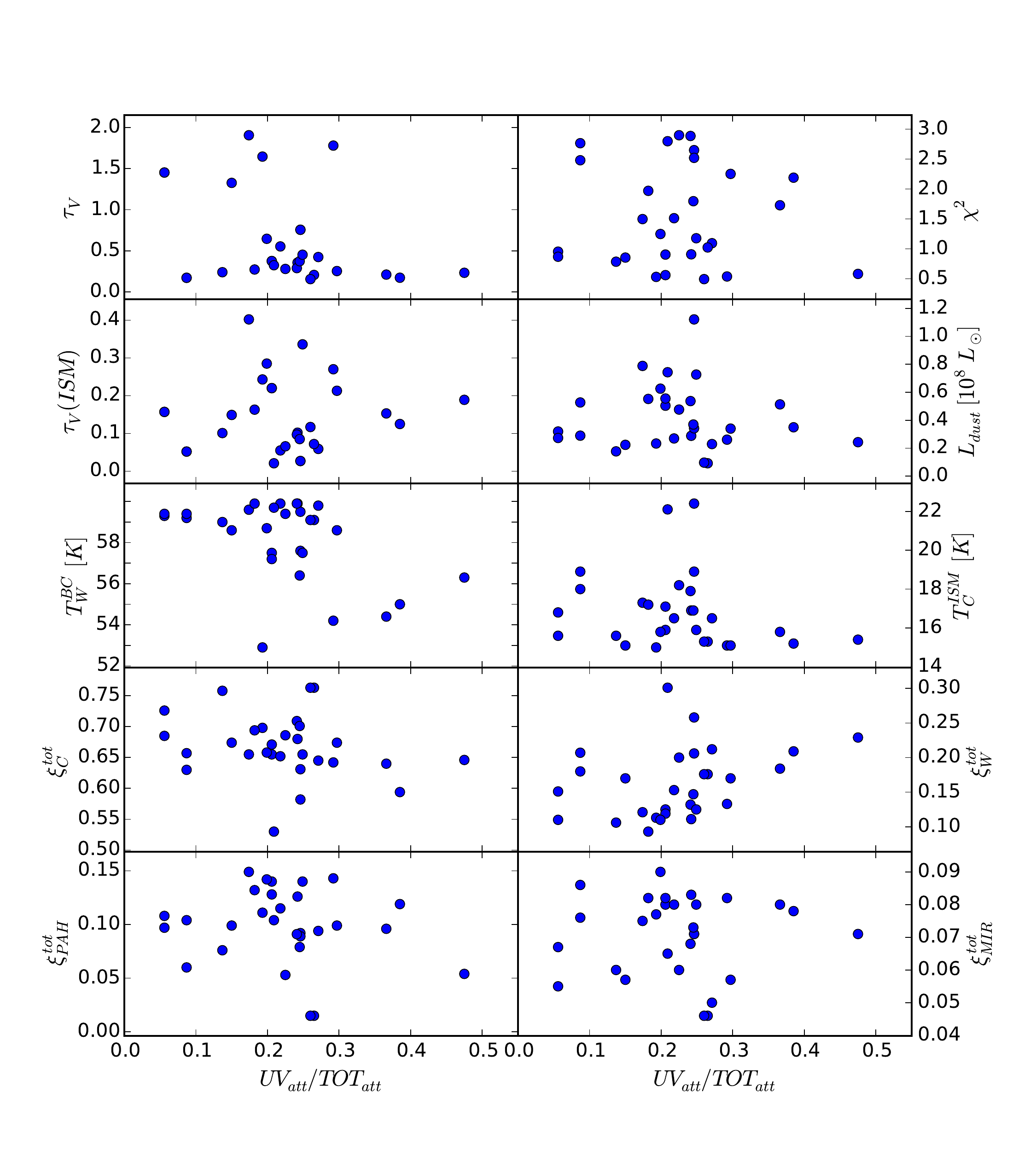}
\caption{{\tt MAGPHYS} output parameters plotted against $UV_{att}/TOT_{att}$: $\tau_V$ (the total effective V-band absorption optical depth of the dust), $\chi^2$ of the best fit model, $\tau_V$ (the effective V-band absorption optical depth of the dust in the ambient ISM), $L_{dust}$ (the dust luminosity), $T_W^{BC}$ (the equilibrium temperature of warm dust in stellar birth clouds), $T_C^{ISM}$ (the equilibrium temperature of cold dust in the ambient ISM), $\xi_C^{tot}$, $\xi_W^{tot}$, $\xi_{PAH}^{tot}$, $\xi_{MIR}^{tot}$ (the global contributions by cold and warm dust, PAH and hot mid-infrared continuum  components, respectively, including stellar birth clouds and the ambient ISM, to the total infrared luminosity of a galaxy).}
\label{fig:mag_UV_TOT_2}
\end{center}
\end{figure*}

%===================================================================
%===================================================================

\section*{Acknowledgments}
The authors thank E. Tempel and collaborators for sharing the SDSS data with us. 
The authors thank J. D. Smith, H.-W. Rix and F. Walter for helpful conversations  in the course of this project.
M. J. K. thank M. \L{}yczek, M. Cluver and Ch. Magoulas for the support.
The authors would also like to thank the anonymous referee for providing us with very constructive comments.
We thank J. D. Smith for providing an early version of his paper for comparison.
During the course of this research, M. J. K. received funding support from the DLR through Grant 50 OR 1115 and the National Research Foundation (NRF; South Africa). 
K. S. acknowledges funding from a Marie Curie International Incoming Fellowship.
B.G. gratefully acknowledges the support of the Australian Research Council as the recipient of a Future Fellowship (FT140101202).
S.C.O.G. acknowledges financial support from the Deutsche Forschungsgemeinschaft  via SFB 881, ``The Milky Way System'' (sub-projects B1, B2 and B8) and SPP 1573, ``Physics of the Interstellar Medium'' (grant number GL 668/2-1), and by the European Research Council under the European Community's Seventh Framework Programme (FP7/2007-2013) via the ERC Advanced Grant STARLIGHT (project number 339177).
This research made use of (1) Montage, funded by the National Aeronautics and Space Administration’s Earth Science Technology Office, Computation Technologies Project, under Cooperative Agreement Number NCC5-626 between NASA and the California Institute of Technology. Montage is maintained by the NASA/IPAC Infrared Science Archive.
(2) the NASA/IPAC Extragalactic Database (NED) which is operated by the Jet Propulsion Laboratory, California Institute of Technology, under contract with the National Aeronautics and Space Administration. This research has made use of NASA’s Astrophysics Data System Bibliographic Services. PACS has been developed by a consortium of institutes led by MPE (Germany) and including UVIE (Austria); KU Leuven, CSL, IMEC (Belgium); CEA, LAM (France); MPIA (Germany); INAF-IFSI/OAA/OAP/OAT, LENS, SISSA (Italy); IAC (Spain). This development has been supported by the funding agencies BMVIT (Austria), ESA-PRODEX (Belgium), CEA/CNES (France), DLR (Germany), ASI/INAF (Italy), and CICYT/MCYT (Spain).

\bibliographystyle{apj}
\bibliography{biblio}

\end{document}